\documentclass[lettersize,journal]{IEEEtran}
\usepackage{amsmath,amsfonts}
\usepackage{algorithmic}
\usepackage{algorithm}
\usepackage{array}
\usepackage[caption=false,font=normalsize,labelfont=sf,textfont=sf]{subfig}
\usepackage{textcomp}
\usepackage{stfloats}
\usepackage{url}
\usepackage{verbatim}
\usepackage{graphicx}
\usepackage[colorlinks=true, allcolors=blue]{hyperref}
\usepackage{cite}
\usepackage{hyperref}

\begin{document}

\title{Bending beams for 6G near-field communications}

\author{{\IEEEauthorblockN{Sotiris Droulias, Giorgos Stratidakis,~\IEEEmembership{Member, IEEE}, and Angeliki Alexiou,~\IEEEmembership{Member,~IEEE}} }
\thanks{The authors are with the Department of Digital Systems, University of Piraeus, Piraeus 18534, Greece (Corresponding author: Sotiris Droulias, e-mail: sdroulias@unipi.gr)}}

\maketitle

\begin{abstract}
Future wireless connectivity is envisioned to accommodate functionalities far beyond broadband data transmission over point-to-point direct links, enabling novel scenarios, such as communication behind blockers and around corners, and innovative concepts, such as situational awareness, localization and joint communications and sensing. In this landscape, beams that are able to propagate on bent paths are ideal candidates for dynamic blockage avoidance, interference management in selected regions, and user connectivity on curved trajectories. In this work, we study beam shaping for applications in near-field wireless connectivity. We explain the underlying mechanism of beam bending and we present the design principles for tailoring the curvature of the propagation trajectory. We discuss design aspects for generation of such beams with large arrays and analyze the impact of several parameters on their performance, including the beam's footprint shape, the aperture size, the inter-element spacing, the sub-array selection of active elements, the available phase levels of the array elements and the operating frequency. We introduce the concept of near-field virtual routing (NFVR) and we demonstrate that such beams are able to address challenges of high frequency communications, such as dynamic routing, blockage avoidance and energy-efficiency, more efficiently than conventional beamforming.
\end{abstract}


\section{Introduction}
\IEEEPARstart{W}{ireless} communications are nowadays shifting to higher frequencies. Frequency bands, such as the mmWave and THz, are likely to be utilized in the near future, and several works are already exploring the upcoming opportunities \cite{Zhang2020, Linlong2020, Papasotiriou2020, Qian2021, Tang2021, Pan2021, Zhang2021, Basar2021, Stratidakis2021, Khalily2022,Papasotiriou2023}. A direct consequence of frequency upscale is that, for a certain aperture size $D$, a radiating element becomes electrically large, i.e. the ratio $D/\lambda$ increases, due to the smaller wavelength $\lambda$. As a result, the Fraunhofer distance $z_F=2D^2/\lambda$ increases, transferring the near-to-far-field transition to larger distances, thus enabling operation of communications in the near-field of radiating apertures, such as Large Antenna Arrays (LAAs), Reconfigurable Intelligent Surfaces (RISs) and, generally, Large Intelligent Surfaces (LISs) \cite{Singh2022, Eldar2022, Dai2023a, Dai2023b, Hanzo2023, Zhi2023, Kosasih2023, Singh2023, Liu2023, Stratidakis2023b, Droulias2024}. \\
\indent The availability of electrically large surfaces opens up new opportunities for manipulating the wavefront of the radiated wave, to acquire curvature beyond the typical far-field planar form, achieved with conventional beamforming. To date, the interest of wireless communications has been focused on the far-field of antenna systems and, therefore, the concept of wavefront engineering to achieve beams with tailored properties in the near-field has remained mainly within the realm of optics. In the optical frequencies, the concept of self-accelerating waves, i.e. waves that accelerate without the need of an external potential, has motivated the prediction and observation of beams that propagate on curved trajectories without diffracting \cite{Efremidis2019}. For example, shape-preserving propagation on parabolic, elliptic and circular trajectories has been realized with beams of Airy \cite{Siviloglou2007a,Siviloglou2007b}, Mathieu \cite{Zhang2012} and Bessel 
\cite{Kaminer2012} functional form, respectively. Classes of Bessel-like beams \cite{Rosen1995} can follow zigzag-like trajectories that are not necessarily convex. Pin-like beams propagate on a straight line (zero acceleration) with enhanced robustness against turbulence for free-space optical communications \cite{Droulias2023}. \\
\indent Transferring these functionalities in the wireless domain could open up endless possibilities \cite{Chremmos2013,Stratidakis2023,Singh2023}. The capability to flexibly engineer the wavefront, taking into consideration environmental parameters and usage scenario characteristics, opens up new opportunities for a landscape \textit{beyond just communications}, where communication nodes are able to detect the presence of objects, identify the shape of surrounding obstacles, map the environment, focus power \cite{Huang2018, Ahsan2021, Tran2022, Yang2022, Li2023}, localize and track mobility \cite{Wymeersch2020, He2022, Zhang2022, Ma2023}. In a mobile node environment, where the node follows a bent path, beams that reside in the near-field and evolve on curved trajectories can essentially provide near-field virtual routing from the transmitter (TX) to the receiver (RX) without the need of using multiple RISs, i.e. routing on a predefined trajectory that takes place at the physical layer entirely, without requiring intermediate nodes. Tracking mobile RXs with conventional beamforming would require the use of codebooks and multiple beams, therefore sophisticated algorithms and overhead depending on the RX mobility, even if the RX trajectory is known. With bending beams these requirements are relaxed, as a single beam dedicated to the RX path guarantees uninterrupted connectivity, regardless of the RX position and speed. With judicious design, the power delivery to the RX can even be tailored to reach a maximum at desired locations along the RX path. Additionally, while the blocked line-of-sight (LoS) is usually restored with relays or RISs, beams that bend can circumvent obstacles, eliminating the need for intermediate nodes, and minimizing the power requirements. Importantly, in a rich blocker environment, where multiple blockers may change position and dynamically interrupt the TX-RX, TX-RIS and RIS-RX connectivity, a single relay or RIS fails to guarantee uninterrupted connectivity beyond a specific blockage scenario, requiring multiple deployment solutions. Bending beams that can be reconfigured are ideal for dynamic blockage avoidance, minimizing the deployment complexity. The curved nature of these beams is also ideal for interference management applications, where entire areas can be enclosed by virtual boundaries, guaranteeing minimum interference. \\
\indent In this work, we study wavefront engineering for near-field wireless connectivity. We demonstrate how bending beams are able to address the challenges of high frequency communications more efficiently than conventional beamforming, enabling both the adaptation of connectivity to the wireless environment (dynamic blockage avoidance, interference-free regions), as well as the modification of the wireless environment itself according to the desired connectivity (user connectivity on trajectory). The main contributions are summarized as follows. 
\begin{itemize}
    \item The design principles of bending beams and their capabilities are presented and analytically assessed for 6G connectivity.
    \item The application of bending beams in multiple use cases is proposed, including user on trajectory, static/dynamic blockage avoidance, interference-free regions, wireless power transfer/energy efficient connectivity and multi-beam applications.
    \item The concept of near-field virtual routing is introduced for the first time, to the best of the authors' knowledge.
    \item The efficient formation of such beams using large arrays is analyzed with respect to crucial parameters, including the footprint size, the footprint shape, the array's inter-element spacing, the sub-array selection, the available phase levels and the operating frequency.
    \item It is demonstrated that bending beams enable wideband excitation without beam split.
\end{itemize}
\section{Wavefront engineering capabilities}
\noindent Typically, a beam formed at the TX transmits a signal and the RX collects and analyzes the time signature of that signal, which depends on the phase and amplitude of the emitted beam. From a physical layer perspective, a beam can be perceived as a redistribution of phase and power in space. Hence, having control over both the trajectory and the shape of the beam (peak power, beam-width) is essential for wireless applications; a beam that propagates on a bent path with, ideally, invariant features can circumvent blockers, ensuring uninterrupted connectivity with predictable power delivery at any desired location. \\
\indent To express such beams analytically, one seeks for invariant solutions of Maxwell's equations, which for propagation in free-space reduce to the Helmholtz equation ($ \nabla^2+k^2) \textbf{\textrm E}(\textbf{\textrm r})=0$, where $\nabla^2$ is the Laplacian operator, $\textbf{\textrm E}(\textbf{\textrm r})$ is the vector of the electric field at position $\textbf{\textrm r}=(x,y,z)$ and $k=2\pi/\lambda$ is the wavenumber ($\lambda$ is the wavelength). An identical equation holds for the $\textbf{\textrm H}$-field. The Helmoltz equation defines three independent scalar equations, namely one for $E_x$, one for $E_y$, and one for $E_z$. For paraxial waves in the one-dimensional (1D) scalar case, e.g. where the $E$-field is invariant along $y$ and propagates along $z$, the Helmoltz equation takes the form 
\begin{align}
    i\frac{\partial E}{\partial z} + \frac{1}{2k}\frac{\partial^2 E}{\partial x^2} = 0,
    \label{Eq:EqHelmholtz}
\end{align}
where $E$ is the magnitude of the $E$-field. It has been shown \cite{Siviloglou2007a,Siviloglou2007b} that the paraxial wave equation \eqref{Eq:EqHelmholtz} supports an Airy wave solution of the form
\begin{align}
    E(x,z)=E_0 A_i\left[(4\beta k^2)^{1/3} (x-\beta z^2+j\frac{a}{k}z)\right]e^{j\psi(x,z)},
    \label{Eq:EqAiryTHEORY}
\end{align}
where $\psi = \left[2\beta k(x-(2/3)\beta z^2) + a^2/2k\right] -j a(x-2\beta z^2)$, $A_i$ is the Airy function and $E_0$ is a complex constant. The parameters $a$ and $\beta$ are real constants with units of $\mathrm{m}^{-1}$. The input beam footprint, i.e. the beam cross-section at $z=0$, extends on the $xy$-plane and is invariant along $y$. The 1D Airy solution remains propagation-invariant (i.e. diffraction-free) in the transverse plane for an observer that follows the parabolic trajectory $x=\beta z^2$. The parameter $\beta$ controls the curvature of the beam trajectory and is inversely proportional to the beam width. The parameter $\alpha$ controls the beam power through amplitude tapering; note that, for $\alpha=0$, no tapering occurs and the beam has infinite power. The Airy beam belongs to the general class of self-accelerating waves, and is the only diffraction-free solution of the 1D paraxial equation \cite{Efremidis2019}.
%
%
%
\begin{figure}[t!]
\centering
    \includegraphics[width=1\linewidth]{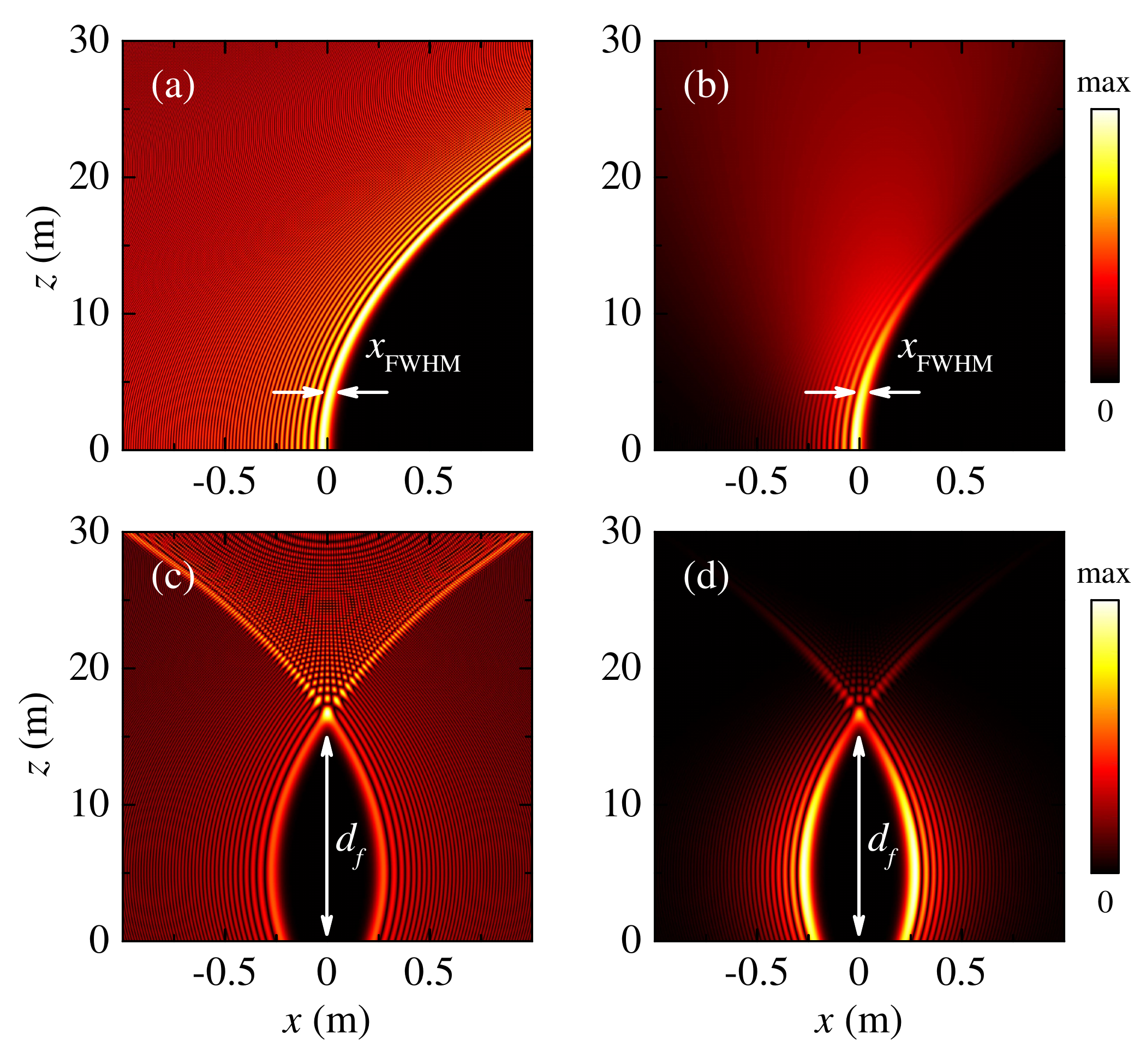}
  	\caption{Capabilities of near-field wavefront engineering. Field amplitude dynamics of ideal 1D Airy beams for (a),(b) propagation  on curved trajectory, and (c),(d) abrupt autofocusing. In (a), (c), the beams have infinite power and in (b),(d) the beams have footprint of finite power (tapered exponentially with $\alpha=4\,\mathrm{m}^{-1}$).}
    	\label{fig:fig01}
\end{figure}
%
%
\subsection{Beam propagation on curved trajectory}
\noindent Using the coordinate transformation $x\rightarrow x_c-x_0,z\rightarrow z_c-z_0$, we can transform \eqref{Eq:EqAiryTHEORY} to evolve on a parabolic trajectory ($x_c,z_c$) with vertex located at ($x_0,z_0$), i.e. along
\begin{align}
    x_c=x_0+\beta (z_c-z_0)^2.
    \label{Eq:EqCAUSTICAiry}
\end{align}
The main lobe of the beam has full width half maximum (FWHM) extent
\begin{align}
    x_\mathrm{FWHM} = \frac{2.278}{(4\beta k^2)^\frac{1}{3}},
    \label{Eq:EqAiryFWHM}
\end{align}
and maximum that is displaced from $x_c$ by
\begin{align}
    \delta x_m = -\frac{1.02}{(4\beta k^2)^\frac{1}{3}}.
    \label{Eq:EqAiryPEAK}
\end{align}
In Fig.\,\ref{fig:fig01}(a) we demonstrate one such beam using \eqref{Eq:EqAiryTHEORY} with $x_0=z_0=0$, $\beta=0.002\,\mathrm{m}^{-1}$ and $\alpha=0$ (infinite aperture). The operation frequency is 150 GHz ($\lambda=2\,\mathrm{mm}$), here and throughout all the examples in this work. 
In Fig.\,\ref{fig:fig01}(b) we plot the same beam with tapered amplitude, using $\alpha=4\,\mathrm{m}^{-1}$. Note that, due to the truncation of the infinite aperture, the main lobe becomes weaker and wider after a certain propagation distance, which decreases with increasing $\alpha$.
\subsection{Autofocusing}
\noindent Airy beams belong to the general class of abruptly autofocusing (AAF) beams \cite{Efremidis2010}, namely beams that are capable of converging and focusing at a selected target distance, while maintaining a low maximum intensity along the entire path propagated from the source. AAF has been investigated for radially symmetric beams and is also possible for 1D beams, with the combination of opposite directed Airy beams, i.e. $E(x,z;x_0,z_0)=E(x-x_0,z-z_0)+E(-x-x_0,z-z_0)$. 
In this case, the power is concentrated at focal distance $d_f$ along the $z$-axis, which can be determined using \eqref{Eq:EqCAUSTICAiry} with $x_c=0,z_c=d_f$ and $x_0\rightarrow x_0+\delta x_m$ [using \eqref{Eq:EqAiryPEAK}], to account for the location of the main lobe's peak with respect to the parabolic trajectory. Solving in terms of $d_f$ leads to
\begin{align}
    d_f = \sqrt{-\frac{x_0+\delta x_m}{\beta}}+z_0.
    \label{Eq:EqAiryFOCAL}
\end{align}
In Fig.\,\ref{fig:fig01}(c), we demonstrate an AAF beam, with $x_0=-0.25\,\mathrm{m}$, $z_0=5\,\mathrm{m}$, $\beta=0.002\,\mathrm{m}^{-1}$ and $\alpha=0$, leading to focusing at distance $d_f=16.7\,\mathrm{m}$. In Fig.\,\ref{fig:fig01}(d), we repeat the same example using $\alpha=4\,\mathrm{m}^{-1}$. Experiments in the optical regime have shown that AAF beams can outperform standard focusing with Gaussians of comparable initial width \cite{Papazoglou2011}. Importantly, they can focus behind potential blockers. Therefore, AAF beams are ideal candidates for applications, including wireless power transfer and energy-efficient communications.
\subsection{Self-healing}
\noindent Self-accelerating waves are known to be able to reconstruct behind obstacles, namely to self-heal \cite{Broky2008}. This is possible, because part from the unblocked beam replaces the blocked part as the beam propagates. Therefore, the self-healing effect is ideal for counteracting blockage.
\section{Tailoring the beam trajectory}
\begin{figure}[t!]
\centering
    \includegraphics[width=1\linewidth]{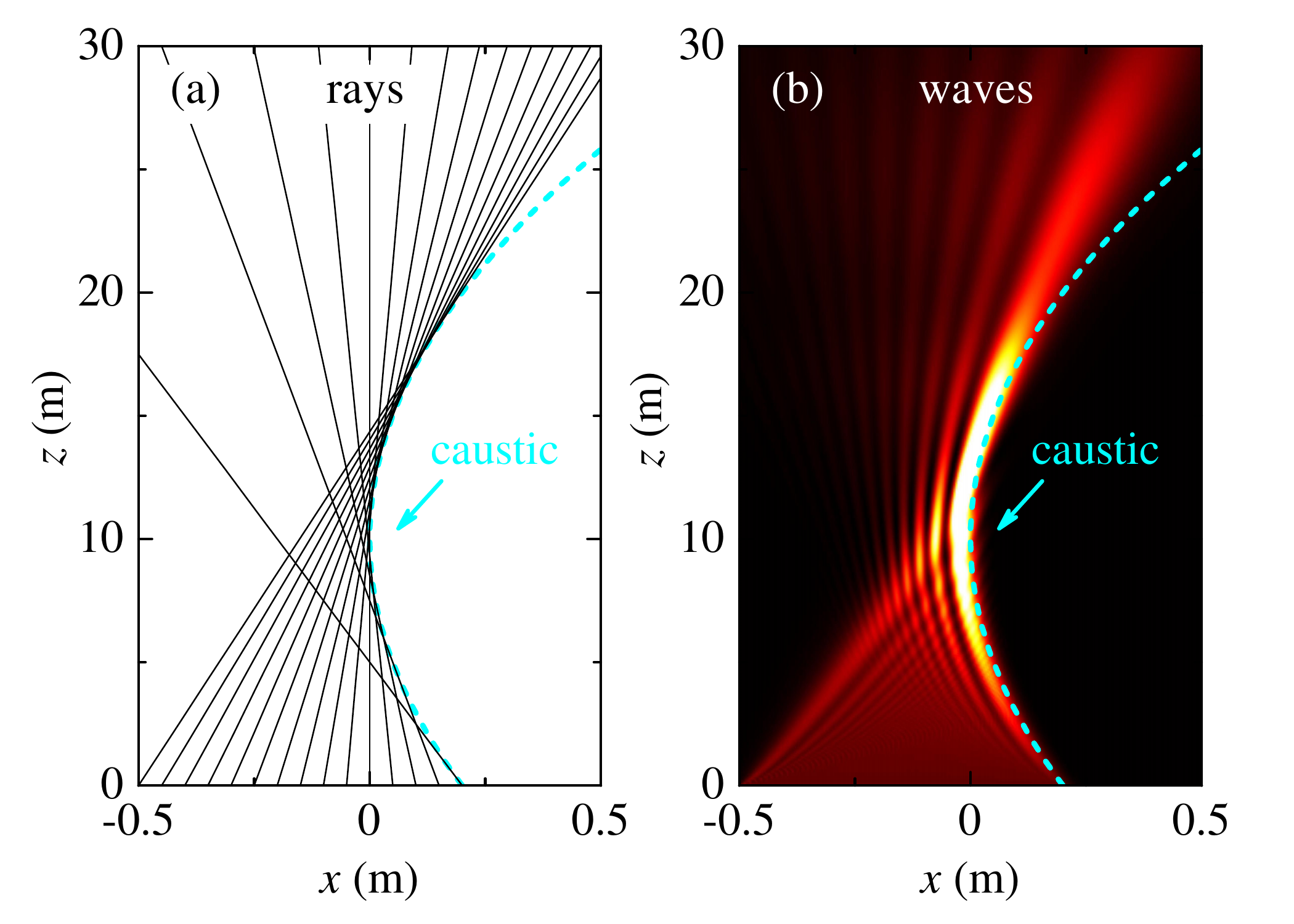}    
	\caption{Near-field beam engineering for propagation along curved trajectory. (a) Ray representation and (b) simulated propagation of equivalent beam. All rays in (a) are tangential to the caustic, which represents the trajectory of the beam's main lobe shown in (b).}
    	\label{fig:fig02}
\end{figure}

\noindent With the analytical description of Airy beams, the impressive capabilities offered by these waves can be directly analyzed, tailored and analytically assessed within the context of relevant applications. However, such ideal beams require specially prepared excitation conditions; their input profile (or footprint) has a spatially varying amplitude and extends to infinity. Even for practical cases where the power must be finite, the footprint follows exponential tapering. In real-life applications the aperture is always finite and, when exciting such beams, it is simpler to have uniform amplitude or amplitude with some conveniently generated profile over the available aperture. Additionally, while such beams propagate strictly on parabolic paths, in many cases it is desirable to be able to launch a beam into trajectories with different curvature profiles. \\
\noindent The design of beams that can follow trajectories of arbitrary functional forms can be easily understood in terms of their equivalent rays \cite{Froehly2011}. While rays necessarily evolve along straight paths, and thus cannot bent, the desired bending results from their envelope, namely their \textit{caustic}, as illustrated in Fig.\,\ref{fig:fig02}(a). The caustic corresponds to the trajectory of the beam's main lobe, which is shown in Fig.\,\ref{fig:fig02}(b) for the equivalent beam. To form the desired caustic, one needs to engineer the slope of the rays at their origin. Given that the slope of the rays is associated with the wavefront curvature of the wave, by engineering the phase of the beam at its starting location, any curved trajectory is possible \cite{Greenfield2011, Froehly2011}. \\
%
%
%
%
%
\indent Let us consider a large aperture array (LAA, RIS, LIS), which generates beams with tailored amplitude and phase characteristics. The aperture is located at the origin of the coordinate system, extends on the $xy$-plane, and generates beams that propagate along the $z$-axis, are invariant along the $y$-direction and bend on the $xz$-plane, as shown in Fig.\,\ref{fig:fig02}. The beam profile at the input plane ($z=0$) has the general form
\begin{align}
    E(x,y,0)=A(x,y) e^{j\phi(x,y)},
    \label{Eq:EqINPUTWAVE}
\end{align}
where $A$ is the amplitude and $\phi$ the phase of the beam's footprint. The caustic of the equivalent rays lies on the $xz$-plane as well, forming a trajectory $x_c=f(z_c)$. The resulting expression between the derivative of the phase $\phi$ at the input plane and the predesigned trajectory $f$ was introduced in \cite{Froehly2011}, and is given by (see Appendix \ref{Sec:AppendixA} for derivation)
\begin{align}
    \frac{d\phi(x)}{dx}=k\frac{df(z_c)/dz_c}{\sqrt{1+(df(z_c)/dz_c)^2}}.
    \label{Eq:EqPHASEnonpar}
\end{align}
For a desired path, solution of \eqref{Eq:EqPHASEnonpar} provides the necessary phase at the input plane, and the trajectory is parametrically expressed in terms of $x$ as \cite{Chremmos2013}
\begin{align}
    (x_c,z_c)=\left(x-\frac{\phi'(x)}{\phi''(x)},-\frac{k}{\phi''(x)} \right).
    \label{Eq:EqCAUSTICpar}
\end{align}
\indent For example, for the parabolic trajectory of \eqref{Eq:EqCAUSTICAiry}, solution of \eqref{Eq:EqPHASEnonpar} in the paraxial regime leads to the input phase
\begin{align}
    \phi(x) = -2\beta k z_0 x - \frac{4}{3}\sqrt{\beta}k\left(\beta z_0^2 -x_0-x\right)^\frac{3}{2},
    \label{Eq:EqAiryPHASE}
\end{align}
corresponding to the Airy beam, which evolves according to \eqref{Eq:EqAiryTHEORY} (see Appendix \ref{Sec:AppendixB} for derivation). 
By choosing $\beta, x_0$ and $z_0$ as input parameters, the phase shifts imposed by \eqref{Eq:EqAiryPHASE} guarantee that the bending beam’s main lobe will evolve on the caustic trajectory $x_c=f(z_c)$ given by \eqref{Eq:EqCAUSTICAiry}. Hence, any user at location with coordinates $x_\mathrm{RX}, z_\mathrm{RX}$ that satisfy the condition $x_\mathrm{RX} = f(z_\mathrm{RX})$ will be reached by the beam. Inversely, the user location can be chosen as an input parameter in Eq.(3), i.e. $x_c = x_\mathrm{RX}$, $z_c = z_\mathrm{RX}$, to determine the necessary $\beta, x_0$ and $z_0$ for the bending beam to reach the specific user. Note that, for $x_0 = z_0 = 0$, solution of \eqref{Eq:EqCAUSTICAiry} yields $\beta = z_\mathrm{RX}^2/x_\mathrm{RX}$, i.e. the parameter $\beta$ that determines the beam curvature is uniquely defined in terms of the user location.
Of course, in the most general case, a user at a certain location can be served by more than one bending beam, because several combinations of the parameters $\beta, x_0$ and $z_0$, can satisfy \eqref{Eq:EqCAUSTICAiry} for a certain choice of $x_\mathrm{RX},z_\mathrm{RX}$. More details on this aspect will be given in Section \ref{Sec:SectionDynBlockAv}. \\
\indent The ray approach conveys a simple and powerful message; the form of the trajectory is primarily determined by the slope of the rays, i.e. the input phase $\phi$ of the wave. Therefore, although curved propagation with invariant characteristics requires engineering of both the amplitude and phase of the beam [see \eqref{Eq:EqAiryTHEORY}], we may relax the amplitude constraints to design beams with reduced complexity. For example, we may use a rectangular aperture of size $L_x \times L_y$ to generate beams with uniform amplitude $A(x,y)$, $|x|<L_x/2,|y|<L_y/2$, as shown in Fig.\,\ref{fig:fig02}(b). In this example, the aperture has dimensions $L_x=L_y=1\,\mathrm{m}$, and the footprint is characterized by $A(x,y)=1\,\mathrm{V/m}$ and $\phi$ given by \eqref{Eq:EqAiryPHASE} with $x_0=0\,\mathrm{m}$, $z_0=10\,\mathrm{m}$ and $\beta=0.002\,\mathrm{m}^{-1}$. We have used the plane wave expansion to numerically propagate the beam from the aperture up to $z=30\,\mathrm{m}$ (see Appendix \ref{AppendixC} for details) and here we plot a cross-section of the beam on the $xz$-plane at $y=0\,\mathrm{m}$. Note how, despite the fact that we have engineered the input phase only, the beam accurately follows the prescribed parabolic trajectory. \\
\indent Importantly, numerous other trajectories can be designed, besides the parabolic, as we will show in the next section. While \eqref{Eq:EqPHASEnonpar} can, in general, be integrated numerically, in many cases the phase profile can be expressed in closed form, e.g. for trajectories of circular, elliptic, power law, and exponential form \cite{Froehly2011, Efremidis2019}. \\
\indent Last, it is important to note that, if desired, it is possible to steer the beam on the transverse $yz$-plane, to compensate possible altitude difference between the TX and the RX. This can be realized by introducing a phase-shift of the form $\phi(x,y) = \phi_x(x) + \phi_y(y)$, where $\phi_x(x)$ tailors the beam bending along the $x$-direction (e.g. as in \eqref{Eq:EqAiryPHASE}) and $\phi_y(y) = k \sin (\theta) y$ steers the beam along the $y$-direction at angle $\theta$, without affecting the bending capabilities of the beam on the $xz$-plane.
\section{User connectivity on curved trajectory}
\noindent The broadcast nature of wireless connectivity has the benefit of serving multiple users without requiring knowledge of their location, at the expense of increased interference and poor efficiency, as the power is distributed in large areas. With directional beamformers it is possible to direct the transmitted signal towards targeted users, thus increasing the gain and reducing the interference. However, because such beams propagate along straight paths, there are certain limitations in the achievable connectivity. In this section we demonstrate how bending beams are able to address the challenges of high frequency communications, such as dynamic routing, blockage avoidance, and energy-efficiency, more efficiently than conventional beamforming. In the following examples, the performance of bending beams is quantified in terms of $|E|^2$, which expresses the local power density of the beam, and, hence, the actual beam propagation also depicts the power distribution in space; the translation of the performance assessment into quantities such as SNR, SINR and Shannon capacity follows directly.
\subsection{User on trajectory and near-field virtual routing}
%
%
\begin{figure}[t!]
\centering
    \includegraphics[width=1\linewidth]{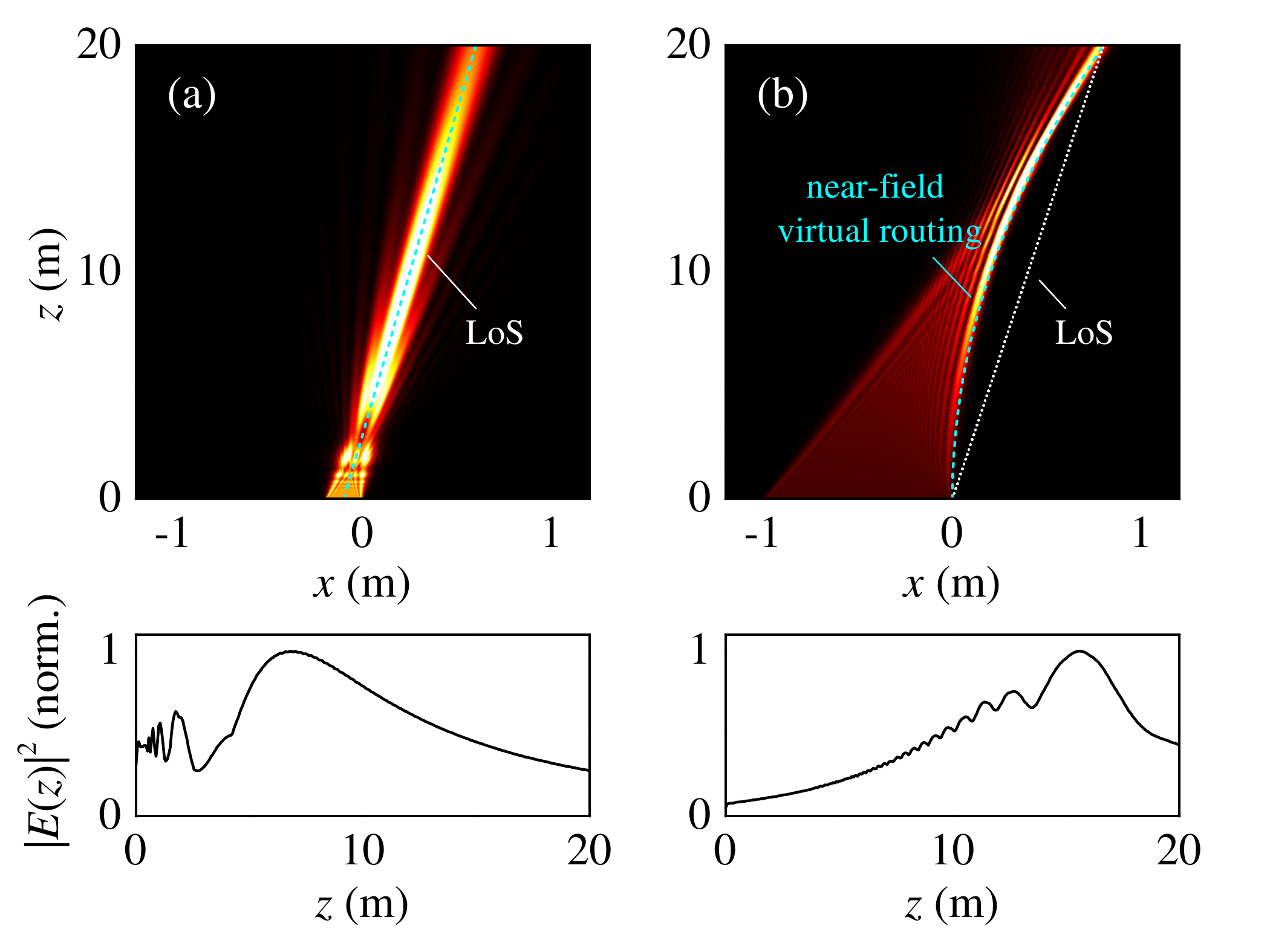}    
	\caption{User connectivity on trajectory. (a) Connectivity on a straight line (LoS) with beam forming. (b) Connectivity on curved trajectory (NFVR) with beam bending. Contrary to conventional beamforming, with bending beams it is possible to achieve connectivity beyond the LoS, and to maximize the power delivered to the RX without increasing the transmitted power.}
    	\label{fig:fig03}
\end{figure}
%
\noindent Beams generated with conventional beamforming propagate along straight paths. Therefore, to ensure connectivity with a mobile user that is moving on a curved trajectory, it is necessary to utilize multiple beams with sophisticated tracking algorithms. Even when there is a priori knowledge of the RX trajectory, unexpected changes in speed and location of the RX might overburden tracking, increasing the overhead. With bending beams, a single beam can be dedicated to a tailored trajectory, ensuring uninterrupted connectivity, regardless of the RX position and mobility. Therefore, bending beams enable connectivity beyond the LoS, essentially creating virtual routing that takes place at the physical layer entirely, without requiring intermediate nodes; we refer to this functionality as near-field virtual routing (NFVR). \\
\indent This concept is demonstrated in Fig.\,\ref{fig:fig03}, where we use an aperture of size $L_x=L_y=0.1\,\mathrm{m}$ to form a conventional beam [Fig.\,\ref{fig:fig03}(a)] and an aperture of size $L_x=L_y=1\,\mathrm{m}$ to form a beam that propagates on a parabolic trajectory with $\beta=0.002\,\mathrm{m}^{-1}$ [Fig.\,\ref{fig:fig03}(b)]. Note that, in the far-field of beams generated with conventional beamforming, the power density decays as $\sim 1/r^2$. Therefore, to increase the signal power to remote users it is necessary to increase the transmitted power. With bending beams it is possible to distribute the transmitted power to selected distances, by tailoring the amplitude characteristics of the footprint. In Fig.\,\ref{fig:fig03}(a), where power is delivered to a remote user via beamforming, the beam enters the far-field after the first few meters and then decays as $\sim 1/r^2$. On the contrary, in Fig.\,\ref{fig:fig03}(b), where the power is delivered to the user via the bending beam, the power gradually increases along the propagation path and is maximized at a distance, where most rays from the aperture converge.
\subsection{Static blockage avoidance and beam resilience}
%
%
%
\begin{figure}[t!]
\centering
    \includegraphics[width=1\linewidth]{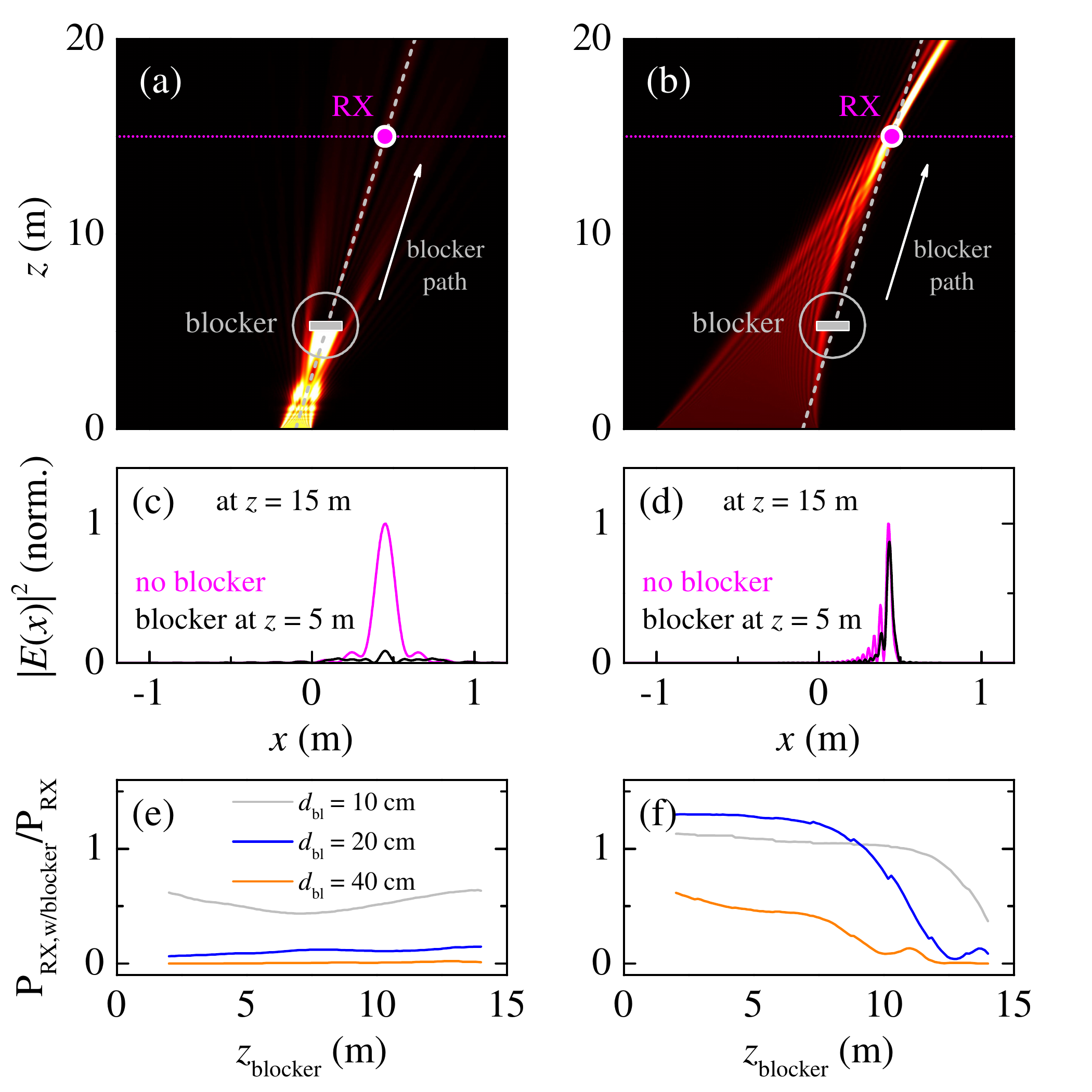}
	\caption{Beam resilience to blockage for (a),(c),(e) conventional beamforming, and (b),(d),(f) beam bending. (a),(b) Numerically propagated beam. (c),(d) Beam cross-section at $z=15\,\mathrm{m}$ with and without the blocker. (e),(f) Ratio of the received power at the RX with and without the blocker, as a function of the blocker position [see dashed lines in (a),(c)] and size. The blocker is an opaque disk on the $xy$-plane with diameter $d_\mathrm{bl}=10\,\mathrm{cm},20\,\mathrm{cm},30\,\mathrm{cm}$.}
    	\label{fig:fig04}
\end{figure}
%
\noindent With conventional beamforming, blockage is usually treated with relays and RISs, which restore the LoS. Relays are active elements and, as such, increase the overall power consumption. RISs are nearly passive, however because the propagation path from the TX to the RX is elongated, the signal to the RX can be severely deteriorated. For a total number $N$ of RISs, the TX-RX path is split into $N+1$ parts of length $d_i$ ($i=1,2,...,N+1$) and the received power decays as $\propto (d_1+d_2+...+d_{N+1})^{-2}$. Taking into account that each RIS usually captures only part of the incident beam, the received power is practically even lower.\\
\indent A straightforward benefit of bending beams is that obstacles along the path from the aperture to the user can be circumvented, without the need for an intermediate node. What is more impressive is that, even if a beam is blocked, it can still outperform conventional beamforming. This is what we demonstrate in Fig.\,\ref{fig:fig04}, where the connection with a RX located at $x_\mathrm{RX}=0.45\,\mathrm{m}$, $z_\mathrm{RX}=15\,\mathrm{m}$ is interrupted by a blocker located at distance $z_\mathrm{blocker}=5\,\mathrm{m}$ along the LoS, from the aperture to the RX. In the examples of Fig.\,\ref{fig:fig04} we use the same parameters as in Fig.\,\ref{fig:fig03} to generate a conventional beam and a bending beam. A cross-section of each beam in the vicinity of the RX is shown in Figs.\,\ref{fig:fig04}(c),(d), with and without the blocker. Clearly, while with conventional beamforming the beam is almost entirely blocked, the bending beam is able to reconstruct (or self-heal) behind the blocker and restore its power with negligible loss. In this example the blocker is an opaque disk on the $xy$-plane with $20\,\mathrm{cm}$ diameter. For other blocker diameters, $d_\mathrm{bl}$, the ratio of the received power with and without the blocker is shown in Figs.\,\ref{fig:fig04}(e),(f), as a function of the blocker position. Note how, with beamforming, the signal is suppressed along the entire blocker path, while bending beams are resilient for most of the blocker's path from the aperture to the RX. In general, smaller blockers are expected to have weaker impact, while larger blockers could possibly interrupt the connection altogether, as already observed in Fig.\,\ref{fig:fig04}(e). While this also holds for bending beams \cite{Inserra2022}, it is possible that the received power may increase with increasing blocker size and even exceed that in the absence of the blocker, as demonstrated in Fig.\,\ref{fig:fig04}(f). This counter-intuitive behaviour results from constructive interference that takes place behind the blocker, due to converging rays that contribute from the side of the blocker, a possibility more challenging to achieve with conventional beamforming, where rays diverge.
\subsection{Dynamic blockage avoidance}
\label{Sec:SectionDynBlockAv}
%
%
%
\begin{figure}[t!]
\centering
    \includegraphics[width=1\linewidth]{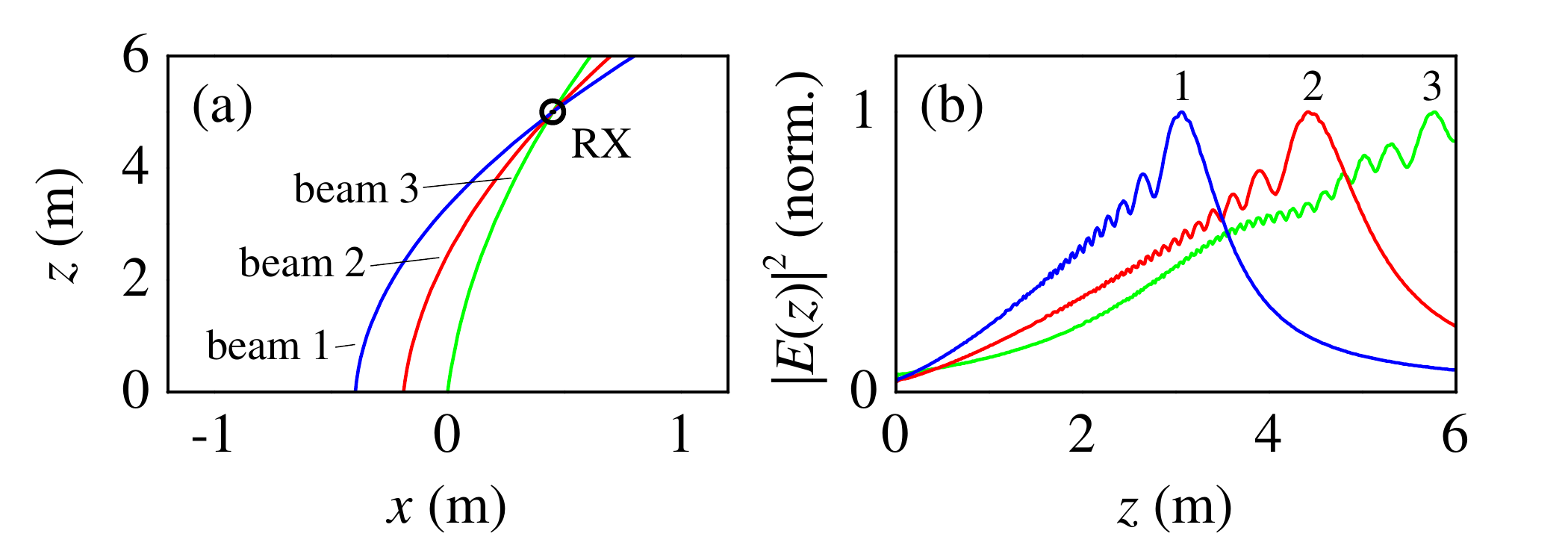}    
	\caption{Dynamic blockage avoidance. (a) The RX located at $x_\mathrm{RX}=0.45\,\mathrm{m},z_\mathrm{RX}=0.5\,\mathrm{m}$ is served by any of the three beams, all generated by the same aperture. The beam is chosen according to the position of the blocker. (b) Normalized peak power along the main lobe of each of the three beams shown in (a).}
    	\label{fig:fig05}
\end{figure}
%
%
\noindent In environments with multiple and/or mobile blockers, bending beams that can be dynamically reconfigured are ideal for dynamic blockage avoidance. Conventional beams propagate on a straight line and, therefore, the path from the aperture to the RX is unique. On the contrary, the curvature of bending beams is tunable and,  therefore, the same target position can be reached via multiple routes, in order to counteract the presence of dynamic blockers. In Fig.\,\ref{fig:fig05} we demonstrate a scenario, where a RX is located at $x_\mathrm{RX}=0.45\,\mathrm{m},z_\mathrm{RX}=0.5\,\mathrm{m}$ and a beam generated by an aperture of size $L_x=L_y=1\,\mathrm{m}$ reaches the RX via three different paths. The chosen beam trajectories illustrated in Fig.\,\ref{fig:fig05}(a) are characterized by $\beta=0.012\,\mathrm{m}^{-1},0.02\,\mathrm{m}^{-1},0.03\,\mathrm{m}^{-1}$ and $x_0=-0.02\,\mathrm{m},-0.2\,\mathrm{m},-0.4\,\mathrm{m}$, respectively. To ensure that the user is reached by all beams, \eqref{Eq:EqCAUSTICAiry} dictates that $z_0=z_\mathrm{RX}-\sqrt{(x_\mathrm{RX}-x_0)/\beta}$ for the each chosen $\beta$, $x_0$ pair. Depending on the position of the blocker, different beams can be dynamically chosen to ensure uninterrupted connectivity, with different power characteristics, as shown in Fig.\,\ref{fig:fig05}(b) for the three example beams.
%
%
\begin{figure}[t!]
\centering
    \includegraphics[width=1\linewidth]{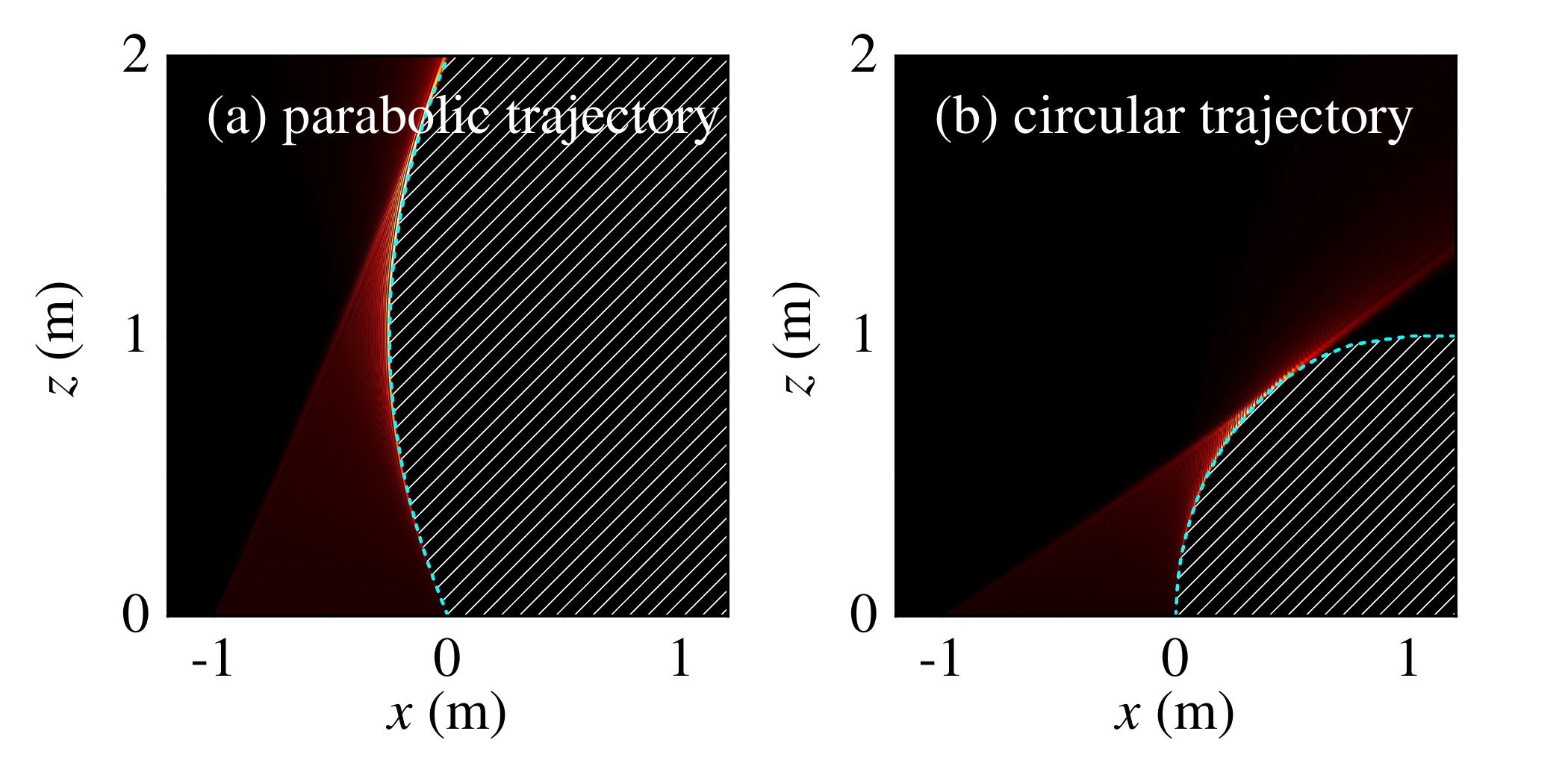}    
 	\caption{Interference-free regions and interference management. The beams decay sharply at the boundary of the shaded areas and, hence, RXs inside these areas are protected from interference. Examples of beams evolving along (a) parabolic and (b) circular trajectories.}
    	\label{fig:fig06}
\end{figure}
%
%
\subsection{Interference-free regions}
\noindent The tailored trajectory in conjunction with the sharp profile of the main lobe is ideal for creating zones or areas, where signals can be controllably suppressed. Contrary to conventional beamforming, these areas can be formed to have curved boundaries, within which the power quickly decays, allowing for the establishment of links practically behind large areas, protected from interference. Depending on the beam curvature, the size of such areas can be chosen at will, even bringing their boundary very close to the aperture. In this case, the beams are characterized by high curvature, and the full form of \eqref{Eq:EqPHASEnonpar} must be taken into account (they fall within the non-paraxial regime). For example, for a parabolic trajectory, the footprint phase takes the form
\begin{align}
    \phi(x) = \frac{1}{4\beta}\left(-2\sqrt{\beta x\left(4\beta x-1\right)} + \mathrm{arcsinh}\left(2\sqrt{-\beta x}\right) \right),
    \label{Eq:EqPHASEnonparPARABOLIC}
\end{align}
which reduces to \eqref{Eq:EqAiryPHASE} for bends of smaller curvature (see Appendix \ref{Sec:AppendixB} for details). In Fig.\,\ref{fig:fig06}(a) we demonstrate an example of one such beam with $x_0=-0.25\,\mathrm{m}$, $z_0=1\,\mathrm{m}$ and $\beta=0.25\,\mathrm{m}$. \\
\indent For beams following a circular trajectory of radius $R$, the necessary phase is calculated as
\begin{align}
    \phi(x) = k R \left(\sqrt{\left(\frac{x}{R}\right)^2-1} - \mathrm{arcsec}\left(\frac{x}{R}\right) \right).
    \label{Eq:EqPHASEnonparCIRCULAR}
\end{align}
In Fig.\,\ref{fig:fig06}(b) we demonstrate an example using a non-paraxial beam of circular trajectory with radius $R=1\,\mathrm{m}$. Due to the finite aperture size, the caustic is successfully formed for the most part of the circular boundary.
\subsection{Wireless power transfer and energy efficient connectivity}
%
%
%
\begin{figure}[t!]
\centering
    \includegraphics[width=1\linewidth]{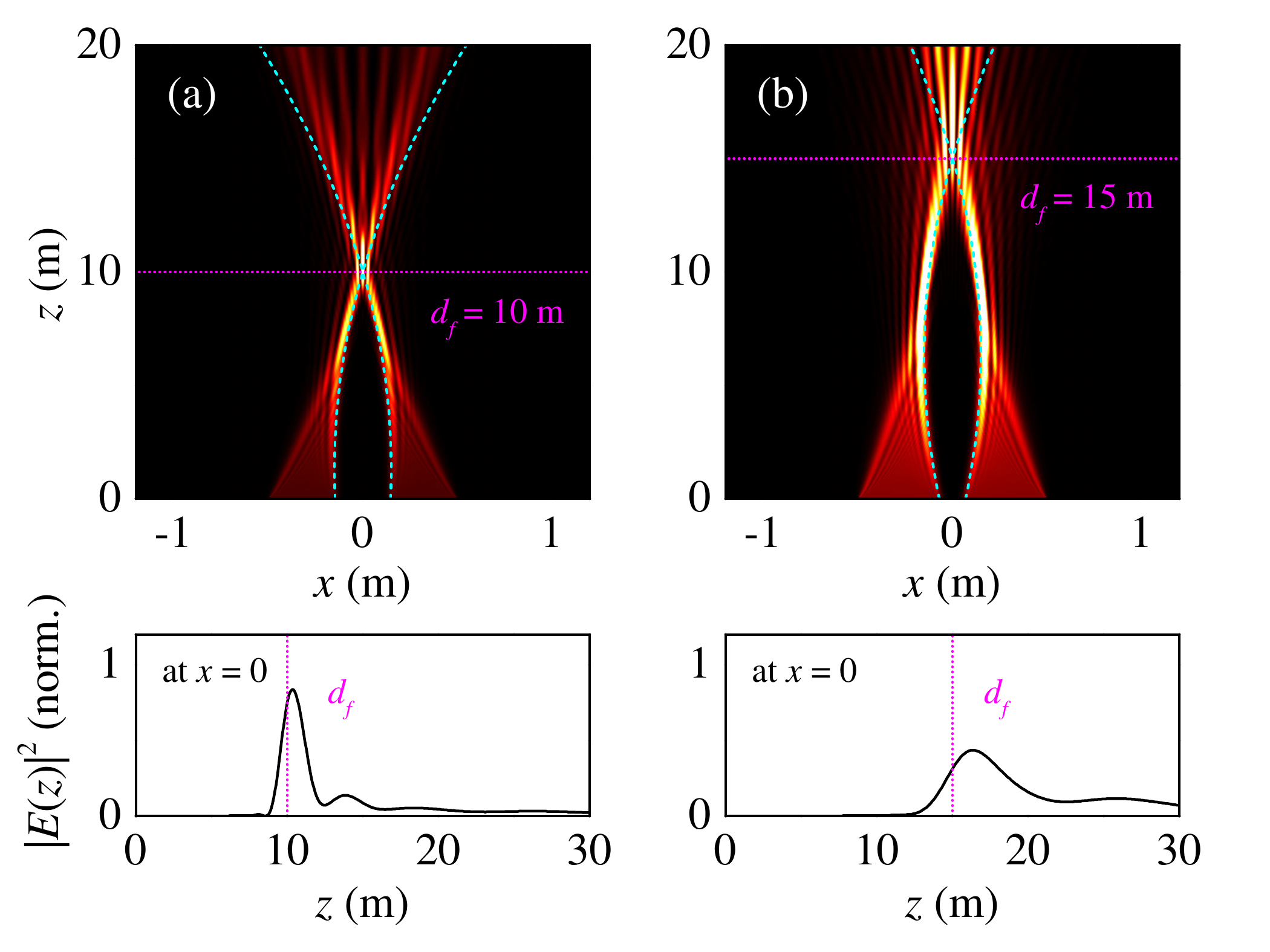}    
 	\caption{Wireless power transfer and energy efficient connectivity. Mirror-symmetric bending beams that propagate on parabolic trajectories are focused at distance (a) $d_f=10\,\mathrm{m}$ and (b) $d_f=15\,\mathrm{m}$. The power abruptly focuses at the desired focal distance, while circumventing possible blockers.}
    	\label{fig:fig07}
\end{figure}
%
%
\noindent The AAF properties of mirror-symmetric bending beams enable the concentration of power at small areas with controllable size, while circumventing possible blockers. Such beams can be used to maximize the power at selected areas and are, therefore, ideal for power transfer applications and energy efficient connectivity: they can be used to amplify the received power without requiring additional power and, inversely, they can be employed with relatively lower power requirements to achieve power levels at remote locations, comparable to those achieved with conventional beamforming. Importantly, due to the relatively small size of the focal area, AAF beams offer high signal-to-interference ratio at the receiver with low transmitted power. \\
\indent In Fig.\,\ref{fig:fig07} we demonstrate the concepts of wireless power transfer and energy efficient connectivity using mirror-symmetric bending beams, generated by an aperture of size $L_x=L_y=1\,\mathrm{m}$. The beams propagate on parabolic trajectories with $\beta=0.002\,\mathrm{m}^{-1}$ and are characterized by $x_0=-0.15\,\mathrm{m}$ and $z_0$ provided by \eqref{Eq:EqAiryFOCAL} with $\delta x_m=0$, using the desired focal distance, which is $d_f=10\,\mathrm{m}$ in Fig.\,\ref{fig:fig07}(a) and $d_f=15\,\mathrm{m}$ in Fig.\,\ref{fig:fig07}(b). Note how, along the $yz$-plane, the power abruptly focuses at the desired focal distance. Because, in each individual beam that comprises the mirror-symmetric AAF beam, the peak of the main lobe is slightly displaced with respect to the trajectory of the caustic [see \eqref{Eq:EqCAUSTICAiry}], the peak of the focal area occurs slightly after $d_f$. It is important to emphasize that this offset is inherent to the functional form of the main lobe, as also predicted by the analytical form of the Airy beam [see \eqref{Eq:EqAiryPEAK}] and, therefore, it is straightforward to take it into account when designing the focal areas.
\subsection{Multi-beam operation}
\begin{figure}[t!]
\centering
    \includegraphics[width=1\linewidth]{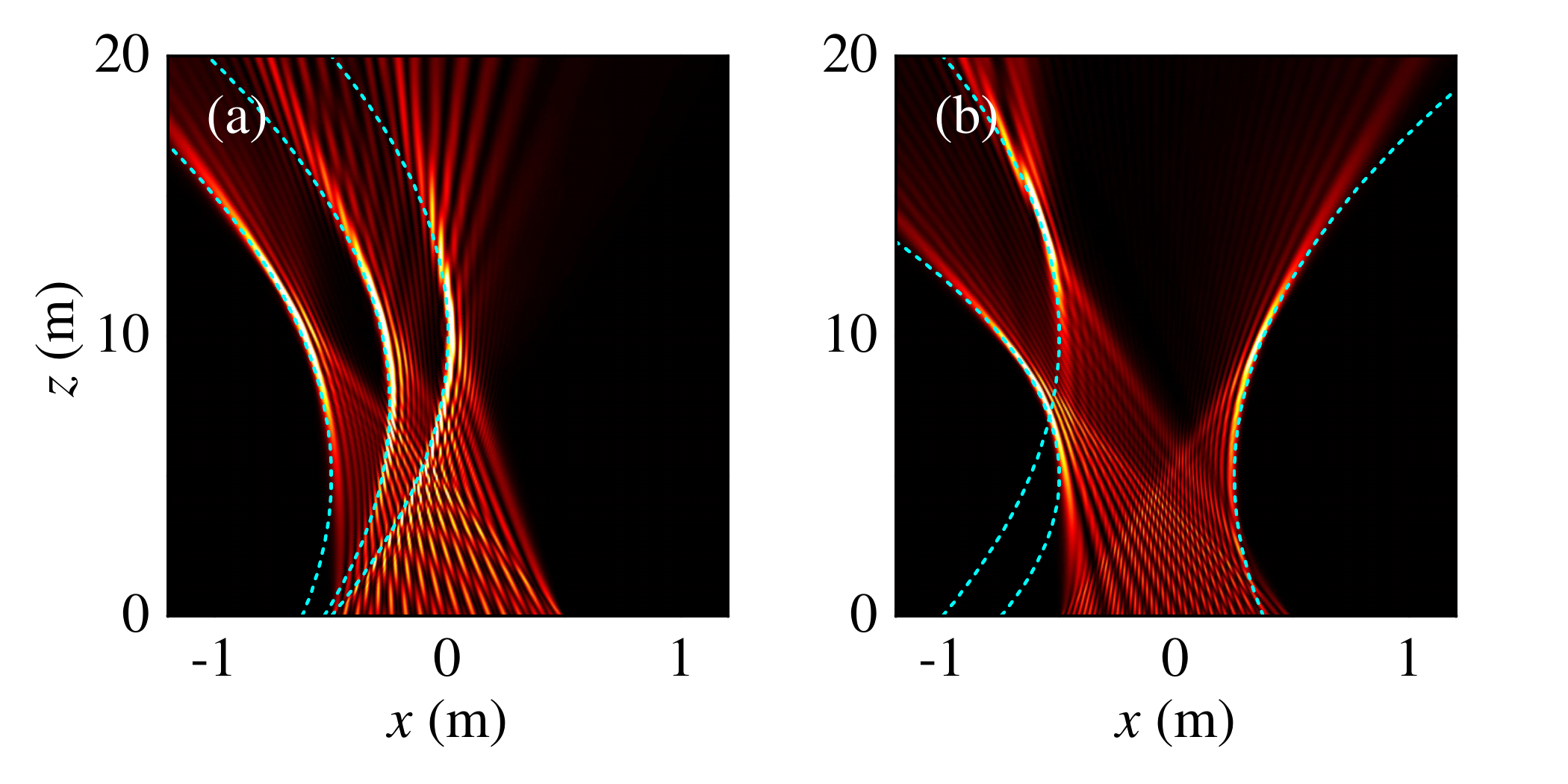}    
	\caption{Multi-beam operation. Three beams are generated to reach RX's located (a) on the same side, (b) on opposite sides of the $x$-axis. The beams can be launched simultaneously in a BDMA scheme or successively in a TDMA scheme.}
    	\label{fig:fig08}
\end{figure}

\noindent For multiple users that reside in the near field of the aperture, the footprint can be split into a multitude of bent beams, to simultaneously serve all users, in a beam division multiple access (BDMA) scenario. In this case, all beams are associated with a common amplitude $A$ and individual phase profiles $\phi_n$, accounting for the $n^\mathrm{th}$ user, which are calculated using \eqref{Eq:EqPHASEnonpar} for each desired trajectory. The footprint is then written as the linear superposition
\begin{align}
    E(x,y,0) = \sum_{n=1}^N w_nA(x,y) e^{j\phi_n(x,y)},
    \label{Eq:EqINPUTWAVEmulti}
\end{align}
which is a generalization of \eqref{Eq:EqINPUTWAVE} to $N$ beams. The weight $w_n$ determines the power distribution among all beams. For beams evolving on a parabolic trajectory, the beam parameters $x_0,z_0,\beta$ acquire distinct values, $x_{0,n},z_{0,n}$ and $\beta_n$, respectively, thus accommodating bent beams, directed towards individual users. \\
\indent As an example, in Fig.\,\ref{fig:fig08} we examine $N=3$ simultaneously generated beams, which evolve along different curved trajectories. In Fig.\,\ref{fig:fig08}(a) all beams have the same curvature $\beta_1=\beta_2=\beta_3=0.005\,\mathrm{m}^{-1}$, and are characterized by the lateral shifts $x_{0,1}=0\,\mathrm{m}, x_{0,2}=-0.25\,\mathrm{m},x_{0,3}=-0.5\,\mathrm{m}$ and the longitudinal shifts $z_{0,1}=10\,\mathrm{m},z_{0,2}=7.5\,\mathrm{m},z_{0,3}=5\,\mathrm{m}$, respectively. All beams propagate towards the same side of the $x$-axis, and with slight adjustments they can be distributed towards any direction, as shown in Fig.\,\ref{fig:fig08}(b). In this example the beams have curvatures $\beta_1=\beta_2=0.005\,\mathrm{m}^{-1}$, $\beta_3=0.01\,\mathrm{m}^{-1}$, and are characterized by the shifts $x_{0,1}=-0.5\,\mathrm{m},x_{0,2}=0.25\,\mathrm{m},x_{0,3}=-0.5\,\mathrm{m}$ and $z_{0,1}=10\,\mathrm{m},z_{0,2}=5\,\mathrm{m},z_{0,3}=5\,\mathrm{m}$, respectively. Note that, this scheme can be as well applied to general multi-user scenarios, in conjunction with a Time Division Multiple Access (TDMA) scheme, where different beams are directed towards individual users at different time slots.
%
%
%
%
%
\section{Design considerations}
\noindent To realize all these fascinating applications it is important to understand the dependencies and trade-offs between the involved parameters, taking into account that LAAs, RISs and general LISs have finite size and are composed of discrete elements with prescribed periodicity. In this section we discuss how to generate bent beams with large arrays and we analyze the efficient formation of such beams with respect to crucial parameters, including the footprint size, the footprint shape, the array's inter-element spacing, the available phase levels and the operating frequency.
\subsection{Bending beam codeword design}
\noindent The formation of bending beams with discrete radiating elements, such as uniform linear arrays (ULAs) and uniform planar arrays (UPAs) requires applying the appropriate time delays between the radiating elements of the array. Such time delays are manifested as phase shifts in the frequency domain and, hence, the input phase distribution $\phi(x)$ provides exactly the phase shifts necessary to generate bending beams with tailored trajectories. \\
\indent Note, however, that the phase $\phi$ of the beam's footprint is a continuous function, while arrays consist of discrete elements periodically distributed in space. Hence, we need to discretize the phase $\phi$ on the array grid, to acquire its values at the position of the array elements, in order to assign the necessary phase shift at each radiator. For a ULA of size $L_x$ that consists of $N_x$ elements, periodically arranged with periodicity $d_x$, this operation translates into expressing $\phi$ at points $x=n_xd_x$, where $n_x\in \mathbb{Z}$. In this case, for the parabolic trajectory \eqref{Eq:EqCAUSTICAiry}, the discretized version of \eqref{Eq:EqAiryPHASE} acquires the form
\begin{align}
    \hat{\phi}(n_x) = -2\beta k d_x z_0 n_x - \frac{4}{3}\sqrt{\beta}k\left(\beta z_0^2 -x_0-d_xn_x\right)^\frac{3}{2},
    \label{Eq:EqAiryPHASEdisc}    
\end{align}
where the hat denotes the discrete version of $\phi$, and
\begin{align}
    n_x = -N_x+1,-N_x+2,\dots,0,
    \label{Eq:EqNx}
\end{align}
i.e. we have considered a ULA extending from $x=-L_x$ to $x=0$.
The steering vector that imposes the phase shifts given by \eqref{Eq:EqAiryPHASEdisc} on the elements of the ULA, takes the form
\begin{align}
\textbf{a}(\beta,x_0,z_0) = \frac{1}{\sqrt{N_x}}\left[e^{j \hat{\phi}(-N_x+1)},e^{j \hat{\phi}(-N_x+2)},\dots, e^{j \hat{\phi}(0)} \right]^T,
    \label{Eq:EqSV}
\end{align}
and the phase discretization can be directly extended to UPAs. With the steering vector parametrized entirely in terms of $\beta,x_0$ and $z_0$, the trajectory of any desired bending beam is directly translated into a respective codeword associated with \eqref{Eq:EqSV}.
\subsection{Impact of footprint size on beam trajectory}
%
%
%
\begin{figure}[t!]
\centering
    \includegraphics[width=1\linewidth]{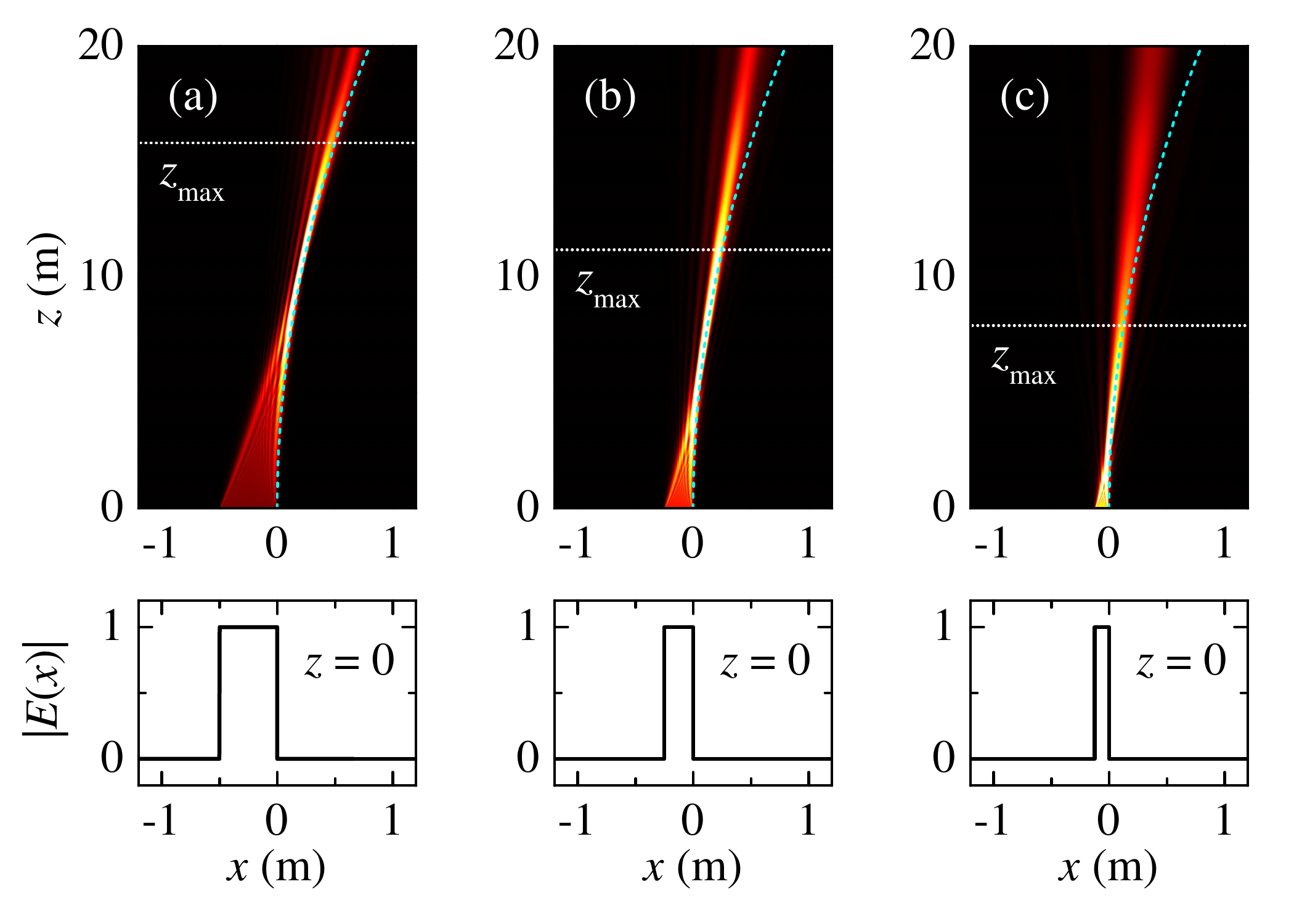}    
	\caption{Impact of footprint size on the beam propagation dynamics. Rectangular aperture of width $L_y=1\,\mathrm{m}$ and length (a) $L_x=0.5\,\mathrm{m}$, (b) $L_x=0.25\,\mathrm{m}$, and (c) $L_x=0.125\,\mathrm{m}$. A cross-section of the footprint is shown in the bottom panels.}
    	\label{fig:fig09}
\end{figure}
%
%
\noindent The size of the footprint plays a crucial role on the maximum distance up to which the beam can propagate along the prescribed trajectory. This is a direct consequence of the fact that all rays are tangential to the trajectory and, hence, smaller footprint involves fewer rays and, consequently, shorter distances before the beam trajectory evolves into a straight path. The size of the footprint can be controlled via the aperture size of the array. For an aperture of size $L_x \times L_y$, it is possible to analytically determine this maximum distance, $z_\mathrm{max}$, which is expressed as (see Appendix \ref{Sec:AppendixD} for details)
\begin{align}
    z_\mathrm{max} = \sqrt{\frac{L_x + \beta z_0^2 + x_0}{\beta}}.
    \label{Eq:EqZmax}
\end{align}
In this expression the aperture extends from $x=-L_x$ to $x=0$. Note that the transverse size $L_y$ plays no role in the bending capability that takes place on the $xz$-plane. In Fig.\,\ref{fig:fig09} we demonstrate the impact of the footprint size on the beam propagation dynamics, for a beam generated by a rectangular aperture with amplitude $A(x,y)=1\,\mathrm{V/m}$ and $\phi$ given by \eqref{Eq:EqAiryPHASE} with $x_0=0$, $z_0=0$ and $\beta=0.002\,\mathrm{m}^{-1}$. We use $L_y=1\,\mathrm{m}$ and variable $L_x=0.5\,\mathrm{m}$, $0.25\,\mathrm{m}$, $0.125\,\mathrm{m}$, corresponding to $z_\mathrm{max}=15.8\,\mathrm{m},11.2\,\mathrm{m},7.9\,\mathrm{m}$, respectively. In all examples, the numerical simulations verify the analytically predicted distance $z_\mathrm{max}$, which lies well within the aperture's near field. Note that, for the selected aperture dimensions, the Fraunhofer distance $z_F=2L_x^2/\lambda$ becomes $z_F=250\,\mathrm{m},62.5\,\mathrm{m},15.6\,\mathrm{m}$, respectively. In essence, by reducing the size of the aperture, the near-to-far-field transition is brought closer to the array, gradually reducing the ability of a beam to bend and rendering it into a conventional beam that evolves quickly into the far-field, propagating along a straight path. Even for users residing at distances larger than $z_\mathrm{max}$, multiple access is still possible, however taking place on straight paths. Therefore, to gain full control of the beam dynamics, it is necessary to increase $z_\mathrm{max}$, i.e. to either increase the aperture size $L_x$ and/or decrease the beam curvature $\beta$.
\subsection{Impact of footprint shape on beam power distribution}
%
%
%
\begin{figure}[t!]
\centering
    \includegraphics[width=1\linewidth]{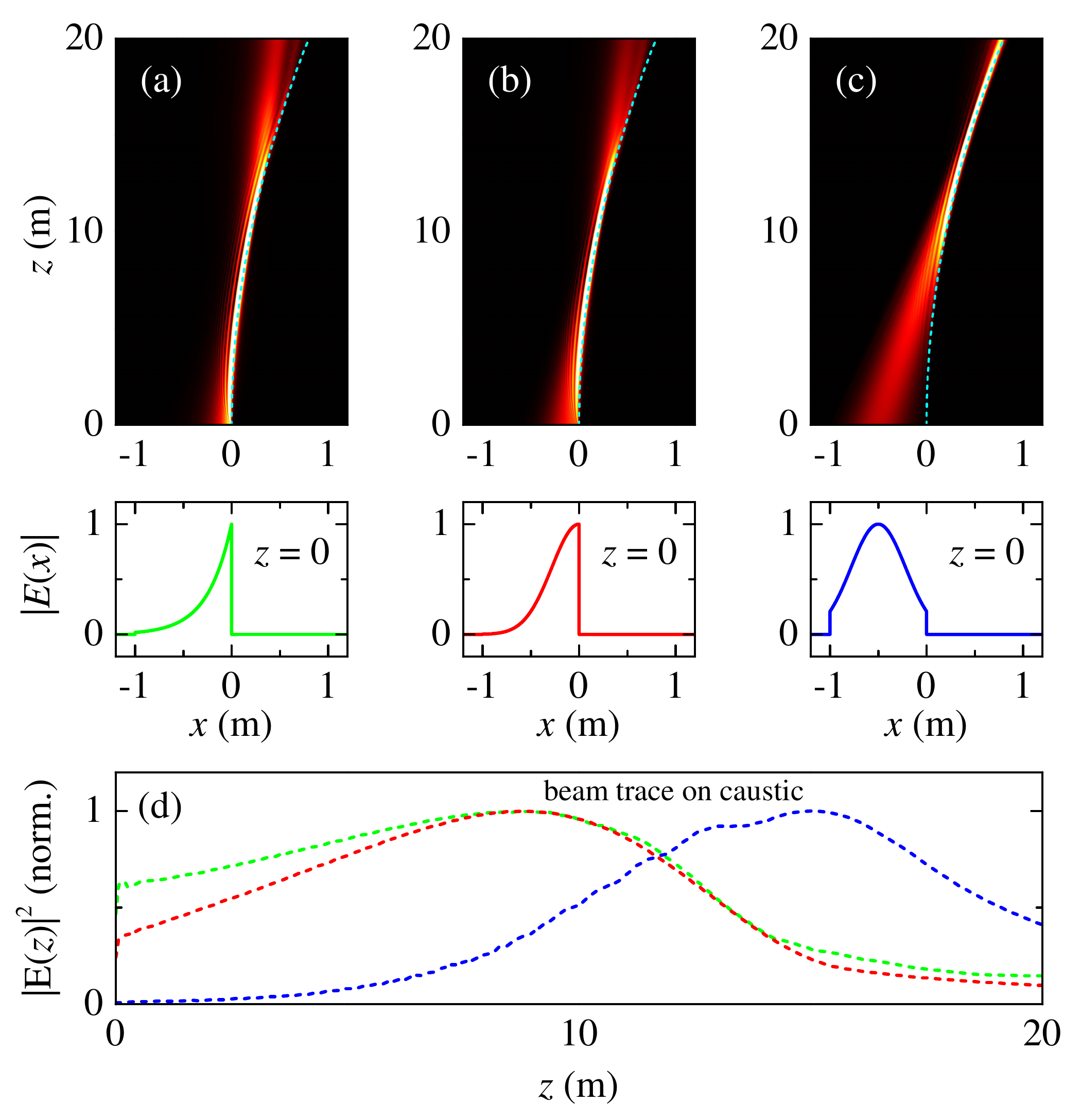}    
	\caption{Impact of footprint shape on the beam propagation dynamics. (a) Exponential tapering. (b) Gaussian tapering centered at $x=0$. (c) Gaussian tapering centered at $x=-0.5\,\mathrm{m}$. A cross-section of the footprint is shown in the bottom panels. (d) Evolution of normalized beam maximum power density along the curved trajectory, for the beams shown in panels (a)-(c).}  
    	\label{fig:fig10}
\end{figure}
%
%
\noindent According to the ray picture, the successful bending of the beam relies primarily on the amount of available rays, i.e. on the spatial extent of the footprint. Its exact shape tunes the power with which each ray contributes along its individual path, to the total caustic. Therefore, the footprint shape controls the local power density of the main lobe along its curved trajectory. In Fig.\,\ref{fig:fig10}(a) we repeat the example of Fig.\,\ref{fig:fig09} using an aperture with $L_x=L_y=1\,\mathrm{m}$, and footprint with exponential tapering, i.e. $A(x,y)\exp(\alpha x)$, with $\alpha=4\,\mathrm{m}^{-1}$. Next, we use Gaussian tapering, i.e. $A(x,y)\exp(-(x-x_m)/\sigma^2)$, with $\sigma=0.4\,\mathrm{m}$, centered at $x_m=0$ in Fig.\,\ref{fig:fig10}(b), and at $x_m=-0.5\,\mathrm{m}$ in Fig.\,\ref{fig:fig10}(c). By transferring the maximum of the footprint's amplitude towards the negative $x$ side, we promote rays that contribute to farther distances. This is evident in the power density of the main lobe along the beam trajectory, which is demonstrated in Fig.\,\ref{fig:fig10}(d) for the three beams. Therefore, we can tailor the footprint's phase to design the beam path and, independently, we can engineer the footprint's amplitude, to control the power delivery at desired distances.
\subsection{Requirements for inter-element spacing}
\noindent To generate beams using large aperture arrays it is necessary to discretize the continuous footprint, i.e. to sample the amplitude and phase of the footprint in discrete steps, which are determined by the periodicity $d_\mathrm{UC}$ of the unit cells in the array. The smaller the inter-element spacing $d_\mathrm{UC}$, the more densely the footprint is sampled and, therefore, rapid spatial variations of the footprint over the available aperture of the array can be captured more efficiently. Usually, the amplitude changes slowly or is constant, as in our examples. The phase, on the other hand, which is crucial for achieving the desired trajectory, usually changes rapidly, imposing stricter constraints on the spatial separation of the array elements. In fact, the sharper the bending, the larger the $\beta$ and the denser the phase oscillations become (see \eqref{Eq:EqAiryPHASE}), naturally raising concerns about limitations in the minimum achievable bending curvature. \\
%
%
%
\begin{figure}[t!]
\centering
    \includegraphics[width=1\linewidth]{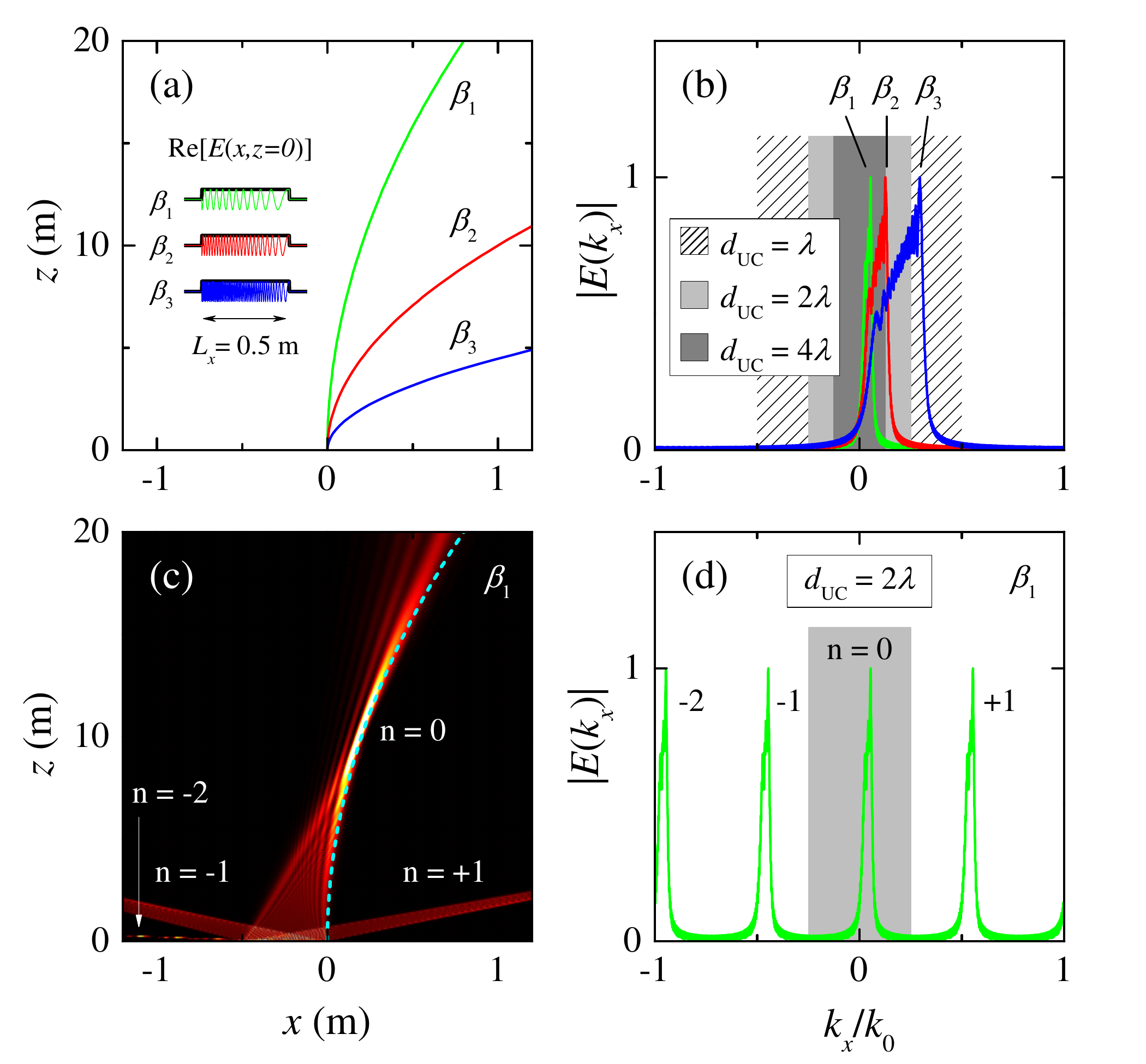}    
	\caption{Impact of inter-element spacing and associated footprint sampling on propagation dynamics. (a) Beam trajectory tailoring with $\beta_1 = 0.002\,\mathrm{m}^{-1}$, $\beta_2 = 0.01\,\mathrm{m}^{-1}$, $\beta_3 = 0.05\,\mathrm{m}^{-1}$. The inset shows the phase oscillations of the footprint. (b) $k$-content of beams shown in (a). The shaded areas denote the maximum inter-element spacing required to adequately sample each beam. (c) Propagation of beam with $\beta_1$, using an array with $d_\mathrm{UC}=2\lambda$ (undersampling). (d) $k$-content of beam shown in (c).}
    	\label{fig:fig11}
\end{figure}
%
%
\indent According to the sampling theorem in space \cite{Orfanidis2016}, the Nyquist criterion requires that the spatial sampling step must be $d_\mathrm{UC}<\pi/B$, where $B=\mathrm{max}(|k_x|)$ is the maximum $k$-component or the spatial bandwidth. To acquire an estimate of the minimum spatial sampling rate for parabolic trajectories, let us calculate the bandwidth of the Airy beam. The power spectrum of the Airy beam with exponential tapering given by \eqref{Eq:EqAiryTHEORY} can be calculated in closed form, and is Gaussian with bandwidth (see Appendix \ref{Sec:AppendixE} for details)
\begin{align}
    \frac{\delta k_\mathrm{FWHM}}{k} = 2\sqrt{\ln2\frac{2\beta}{a}}.
    \label{Eq:EqBW}
\end{align}
For the examples of Fig.\,\ref{fig:fig01}, where $\beta=0.002\,\mathrm{m}^{-1}$ and $\alpha=4\,\mathrm{m}^{-1}$, we find that the sampling criterion is satisfied with $d_\mathrm{UC}<9.5\lambda$, allowing for sparsely arranged unit cells. Besides reducing the design complexity of the array, sparse inter-element separation suppresses the interaction between neighboring unit cells, thus minimizing mutual coupling; each element can be treated as being isolated. Of course, undersampling ($d_\mathrm{UC}>\lambda/2$) gives rise to diffraction orders, i.e. to secondary (non-broadside) caustics that propagate at angles $\mathrm{sin}^{-1}(m \lambda/d_\mathrm{UC})$, $m=\pm1,\pm2,\dots$, which can be undesired. From \eqref{Eq:EqBW} it is evident that, with increasing $\beta$, the bandwidth increases, in turn imposing conditions for denser sampling on sharper trajectory bents.\\
\indent To gain some further insight, in Fig.\,\ref{fig:fig11} we repeat the example of Fig.\,\ref{fig:fig09}(a) for different values of the parameter $\beta$, namely $\beta_1=0.002\,\mathrm{m}^{-1}$, $\beta_2=0.01\,\mathrm{m}^{-1}$ and $\beta_3=0.05\,\mathrm{m}^{-1}$. The beam trajectory for each case is depicted in Fig.\,\ref{fig:fig11}(a), and a cross-section of the footprint is also shown in the inset. With increasing $\beta$, necessary to impose sharper bents, the phase oscillations change more rapidly, in turn leading to an expanded $k$-content, as shown in Fig.\,\ref{fig:fig11}(b). The shaded areas denote the maximum $d_\mathrm{UC}$ required to adequately sample the footprint with $k$-content contained within each region. Note that, while $d_\mathrm{UC}=\lambda/2$ always suffices to sample all footprints completely, for $\beta_1$ sampling can be performed successfully even with $d_\mathrm{UC}\approx 4\lambda$. This results in the excitation of secondary beams, as shown in Fig.\,\ref{fig:fig11}(c), where the beam characterized by $\beta_1$ is sampled with $d_\mathrm{UC}=2\lambda$. In Fig.\,\ref{fig:fig11}(d) we present the $k$-content of the same beam, where the effect of undersampling is clearly seen: the total power is equally split between all beams and, hence, for a total of $N$ diffracted beams, a fraction of $1/N$ of the total power remains in the desired diffraction order ($n=0$).
\subsection{Beam excitation with limited number of elements}
%
%
%
\begin{figure}[t!]
\centering
    \includegraphics[width=1\linewidth]{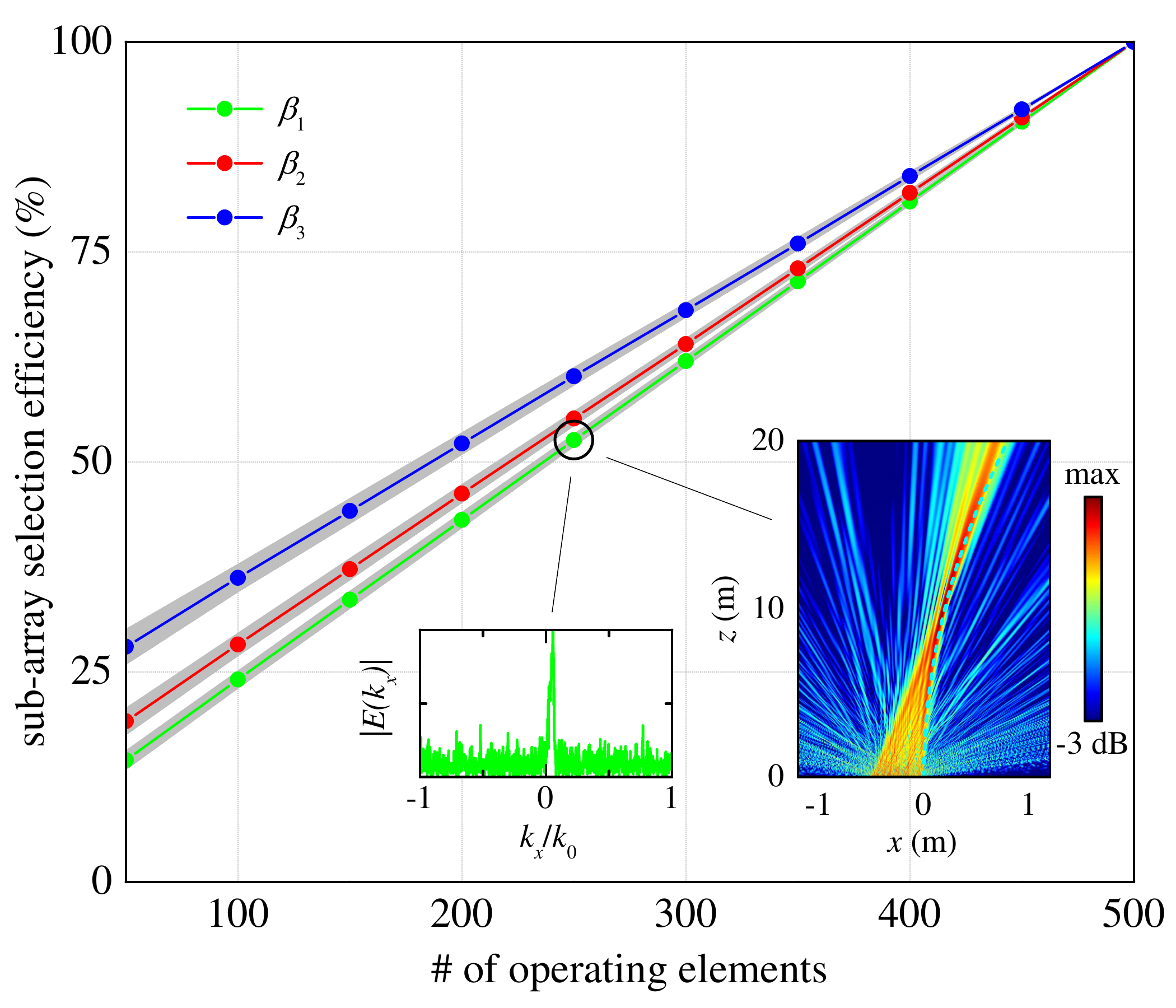}
	\caption{Sub-array selection efficiency for beams generated with limited number of elements. The beam parameters are $\beta_1=0.002\,\mathrm{m}^{-1}$, $\beta_2=0.01\,\mathrm{m}^{-1}$, $\beta_3=0.05\,\mathrm{m}^{-1}$ and the efficiency is averaged over 100 different realizations for each case. The connected dots represent the average efficiency and the gray zones the standard deviation. \textit{Inset}: example for operation with half number of elements, showing the $k$-content of the beam (left) and the beam propagation (right); the colormap is in log scale, to emphasize the details of the background waves.}
    	\label{fig:figA}
\end{figure}
%
\indent Multiple antenna structures often involve fewer RF chains than the total number of antennas. In this case, there are two aspects to be considered with respect to the performance of bending beams: (a) how many RF chains are available and (b) which group of active elements is selected. Intuitively, the more elements, the more efficiently the beam can be generated. While this is in principle true, as we demonstrated in Fig.\,\ref{fig:fig11}, the beam can be generated with fewer elements at the expense of having undesired sidelobes (diffraction orders). In this example the total aperture size is constant and the spacing between individual elements is increased. What we observe in Figs.\,\ref{fig:fig11}(c),(d) is actually the performance with half as much elements. Interestingly, the beams can be still generated if some elements at random positions are not operating.  This has the effect of distributing the power that is lost from the beam towards random directions, contrary to the previous case, where it is directed towards well-defined diffraction orders. \\
\indent To quantify the impact of operation with limited number of elements, we define the sub-array selection efficiency as the ratio of the power that remains on the bending beam over the total power spent to drive the specific sub-array selection. In Fig.\,\ref{fig:figA} we calculate the average efficiency over 100 different realizations, where the non-operating elements are at random positions along the $x$-axis in the array. As expected, the average efficiency reduces with less operating elements. Interestingly, compared to our findings in Fig.\,\ref{fig:fig11}, the efficiency is now slightly higher. This is no surprise; due to the random positions of the operating antennas, undesired waves are generated towards random directions (see inset), some of which along the bending beam path, effectively returning some of the lost power to the beam. Importantly, the standard deviation (shown as gray zones) is only a small fraction of the average efficiency. Consequently, any sub-array selection of active elements ensures similar efficiency.
\subsection{Beam footprint sampling with quantized phase levels}
%
%
%
\begin{figure}[t!]
\centering
    \includegraphics[width=1\linewidth]{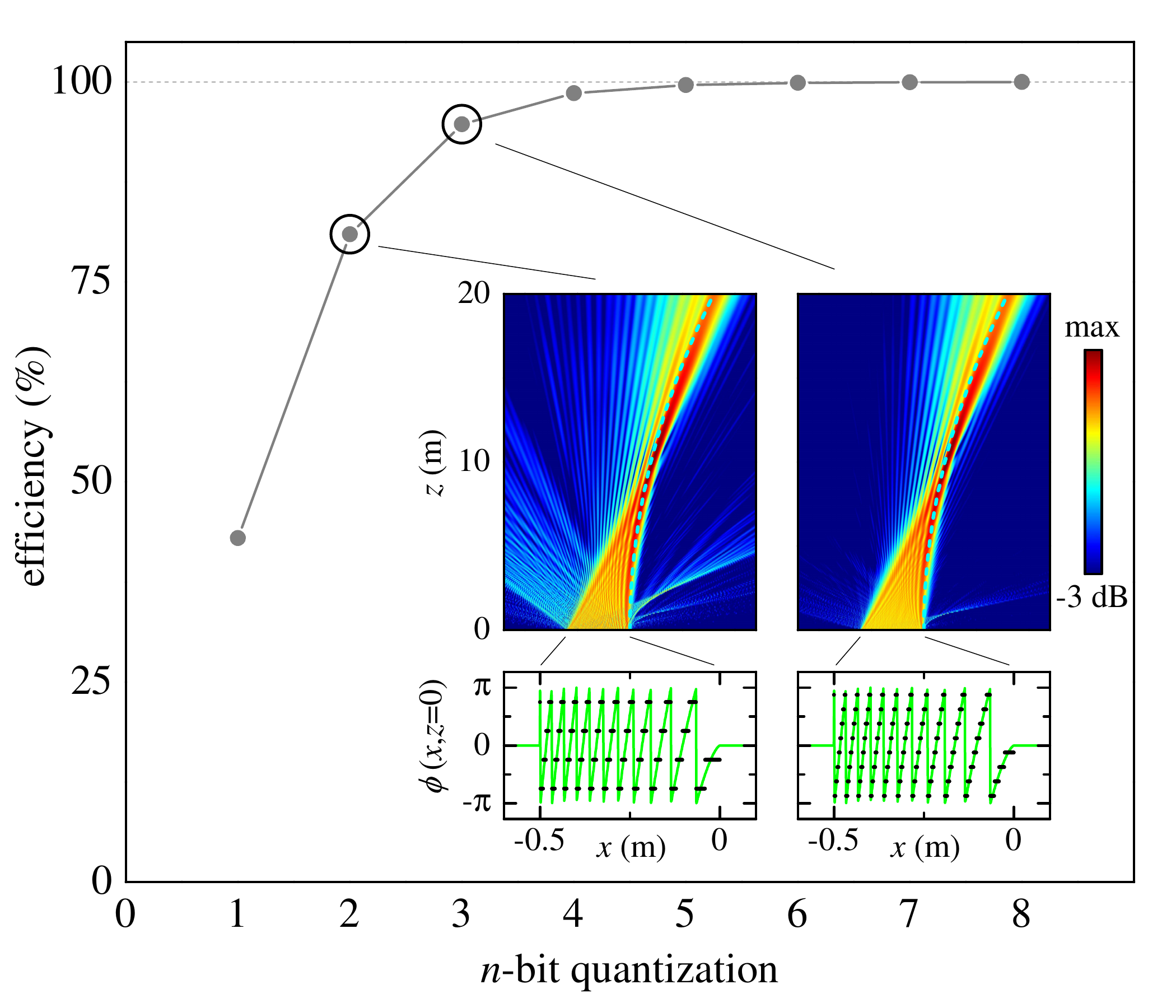}
	\caption{Phase quantization efficiency for beams with $\beta=0.002\,\mathrm{m}^{-1}$, generated by a $n$-bit array. \textit{Inset}: propagation dynamics for $n = 2$ and $n = 3$; the colormaps are in log scale, to emphasize the details of the background waves. The beam footprint has uniform amplitude and its phase is discretized in $2^n$ levels (black dots). The green lines are guides to the eye for the full (continuous) phase.}  
    	\label{fig:fig12}
\end{figure}
%
\noindent Typically, in LAAs the phase in each element is directly controlled by the driving currents and, in principle, any value for the phase is feasible. RISs, on the other hand, usually consist of resonant elements and the phase in each element is controlled by individual frequency shifts of the resonances. Therefore, it is usually less demanding to have a prescribed finite set of phases, which the elements can acquire \cite{Pei2021, Ahmed2024}. Sampling the footprint with a discrete set of phase levels has direct consequences on the quality of the beam and the propagation dynamics. \\
\indent To quantify the effect of phase quantization, we define the phase quantization efficiency as the ratio of the beam power with quantized phase over the beam power with continuous phase. In Fig.\,\ref{fig:fig12} we calculate the phase quantization efficiency for an array with $n$ discrete phase values, which quantizes the $2\pi$ range in $2^n$ levels. We repeat the example of Fig.\,\ref{fig:fig09}(a) with inter-element spacing $d_\mathrm{UC}=\lambda/2$, to eliminate diffracted waves from spatial undersampling and ensure that possible secondary waves result from phase quantization only. Our simulations reveal that phase quantization has the effect of primarily exciting background waves, without severely distorting the beam trajectory, even for 1-bit arrays. As a result, the beam propagates on the prescribed path, however with reduced power, which is scattered to the undesired background waves. The efficiency increases with increasing $n$, as intuitively expected, quickly converging to $\approx100\%$ even with as few as $16$ phase levels ($n=4$). Although the efficiency for a $1$-bit array is $\approx40\%$, it improves quickly as more phase levels are introduced, giving promise for efficient beam generation with $3$-bit or even $2$-bit arrays, as shown in the inset. 
\subsection{Scaling with frequency}
\begin{figure}[t!]
\centering
    \includegraphics[width=1\linewidth]{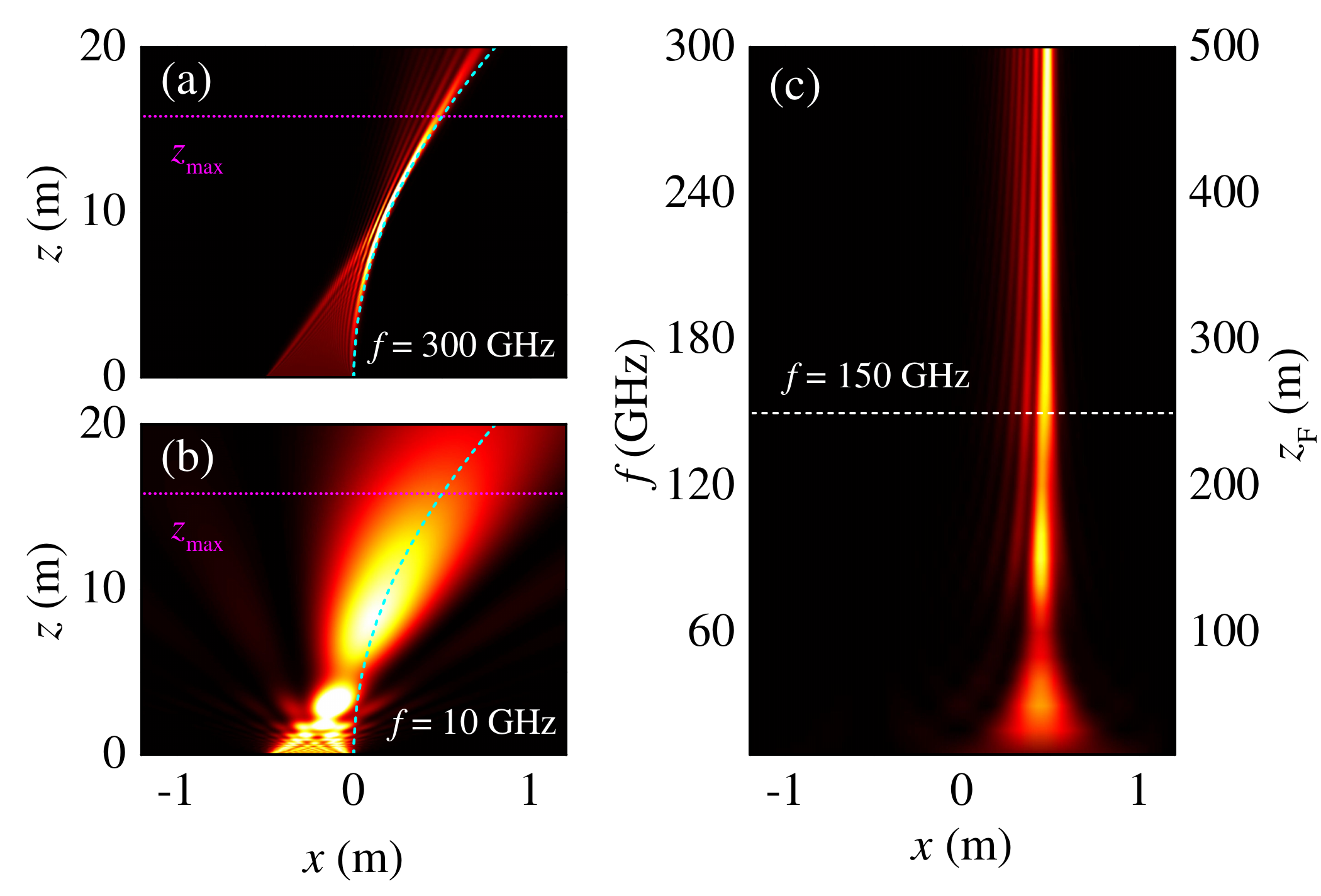}
	\caption{Beam bending efficiency, as a function of the operating frequency. A beam with $\beta=0.002\,\mathrm{m}^{-1}$ is generated by an aperture of size $L_x=L_y=0.5\,\mathrm{m}$ at (a) 300 GHz and (b) 10 GHz. (c) Beam cross-section at $z=z_\mathrm{max}$, as a function of the frequency and the corresponding Fraunhofer distance.}
    	\label{fig:fig13}
\end{figure}

\noindent From the preceding analysis it is evident that efficient beam formation requires a combination of relatively large aperture and adequately sampled footprint. The maximum distance, $z_\mathrm{max}$, up to which the beam can propagate along the prescribed trajectory was previously found to depend primarily on the aperture size. The result of \eqref{Eq:EqZmax} was based on ray analysis, which is a reasonable approximation when the frequency is high enough so that the aperture is electrically large. Hence, it is important to understand how the bending capabilities depend on the operating frequency. \\
\indent In Fig.\,\ref{fig:fig13} we study the bending efficiency as a function of the frequency, using an aperture of size $L_x=L_y=0.5\,\mathrm{m}$ to generate a beam with $x_0=z_0=0$, $\beta=0.002\,\mathrm{m}^{-1}$. Two examples are shown in Figs.\,\ref{fig:fig13}(a),(b), where the operating frequency is set to 300 GHz and 10 GHz, respectively. The dashed magenta line marks $z_\mathrm{max}=15.81\,\mathrm{m}$, the maximum propagation distance predicted by the ray analysis, which is calculated using \eqref{Eq:EqZmax}. The cross-section of the beam at $z=z_\mathrm{max}$ is presented in Fig.\,\ref{fig:fig13}(c), as a function of the frequency and the corresponding Fraunhofer distance, $z_F$, where the white dashed line marks the operation frequency used throughout this work. At the lowest frequency of 10 GHz examined here, the beam formation is poor and, although the beam does bend towards the prescribed trajectory, it resembles a conventional beam, tilted towards the desired angle. This is no surprise, taking into account that the Fraunhofer distance at this frequency is $z_F=16.7\,\mathrm{m}$, which is comparable to $z_\mathrm{max}$, i.e. the beam practically enters its far-field before having the chance to evolve into a bending beam. With increasing frequency the aperture becomes electrically larger, and the distance $z_\mathrm{max}$ is approached more efficiently. This is also evident in the corresponding Fraunhofer distance, which indicates that beam bending takes place deep in the near-field. Note that, once the condition $z_F\gg z_\mathrm{max}$ is fulfilled, the beam is formed efficiently for an extended frequency range. This flexibility is particularly useful for designing beams to operate at different frequencies with the same aperture. It is also important to note that the observed improvement in performance with increasing frequency is due to the fact that the aperture size remains fixed and becomes more and more electrically large, as the wavelength becomes smaller. Alternatively, one could increase the aperture size at lower frequencies to ensure the same performance, independent of the exact operating frequency. This would require keeping the ratio $L_x/\lambda$ constant, which at 10 GHz would lead to impractical size of $L_x = 7.5\,\mathrm{m}$. Hence, higher frequencies offer the benefit of achieving the desired performance with smaller components.

%
%
%
%
\subsection{Wideband excitation without beam split}
\begin{figure}[t!]
\centering
    \includegraphics[width=1\linewidth]{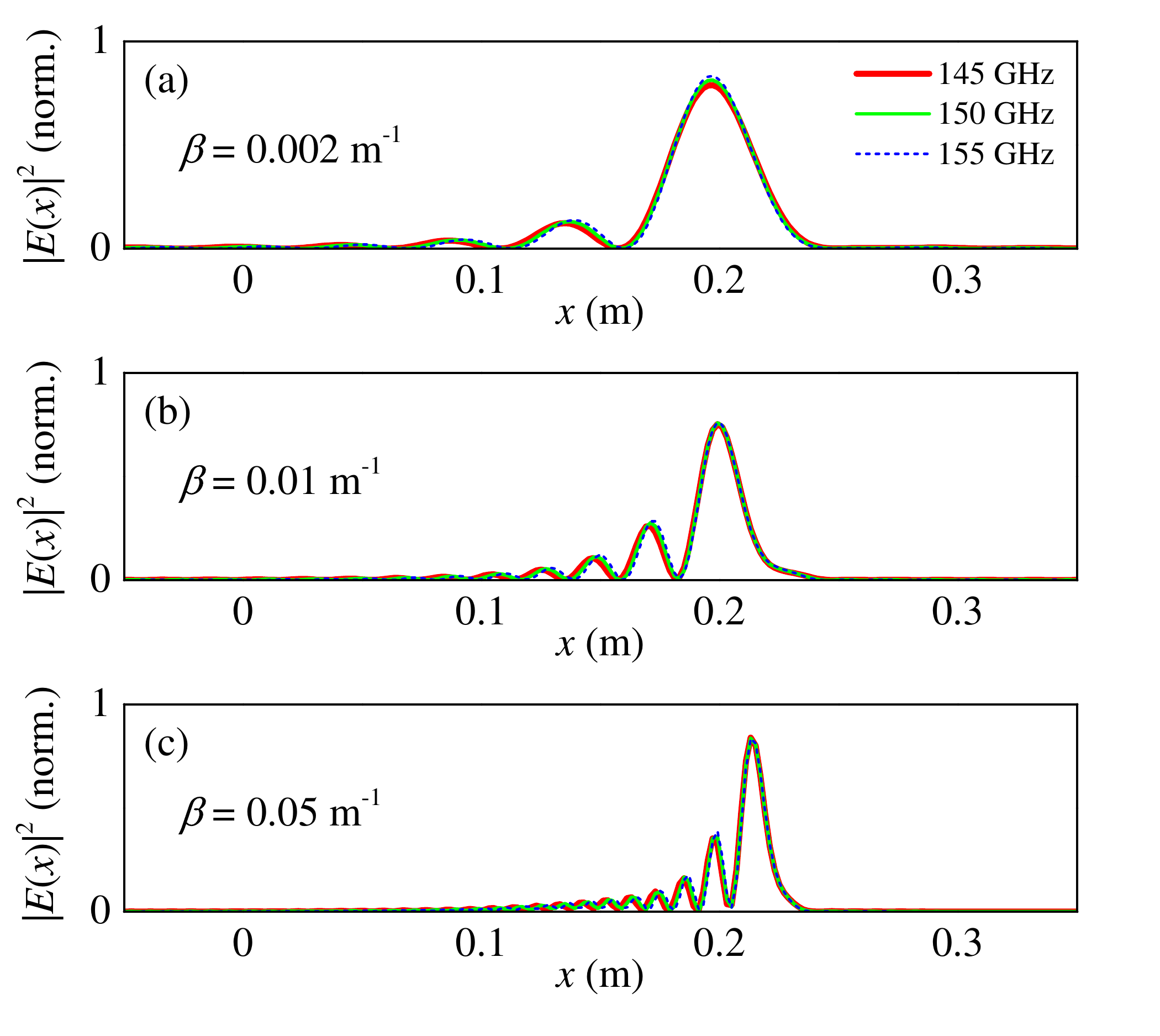}
	\caption{Wideband excitation without beam split. Beam cross-sections at distance $z = z_\mathrm{max}/1.5$ for each beam with (a) $\beta_1=0.002\,\mathrm{m}^{-1}$, (b) $\beta_2=0.01\,\mathrm{m}^{-1}$, and (c) $\beta_3=0.05\,\mathrm{m}^{-1}$. The beams are excited at $145\,\mathrm{GHz},150\,\mathrm{GHz}$ and $155\,\mathrm{GHz}$.}  
    	\label{fig:figB}
\end{figure}

\noindent Data transmission involves time-signals with spectral extent, which is usually a small fraction of the carrier frequency. In such cases, waves can be considered monochromatic to a good approximation, and their performance can be assessed at the carrier frequency (at single $\lambda$). For wideband applications, however, it is possible that the frequency content is so wide that the wavelength distribution that the beam contains cannot be ignored. In this case, the frequency content manifests in space via the wavelength, often leading to the so-called beam split effect \cite{Cui2023}. \\
\indent Because of their special design, bending beams are free from any beam split effect. The reason is that the phase shifts are coordinated to direct the beam's main lobe on the prescribed trajectory, which is independent of the frequency. This fact has already been implied in Fig.\,\ref{fig:fig13}, where the beam cross-sections practically remain invariant beyond a certain frequency, where the ray approximation becomes valid. To better illustrate this property, in Fig.\,\ref{fig:figB} we demonstrate explicit cross-sections of beams excited at different frequencies, namely at $145\,\mathrm{GHz}$, $150\,\mathrm{GHz}$ and $155\,\mathrm{GHz}$. The beams remain practically invariant with frequency, giving promise for wideband applications without beam split.
%
%
%
%
%
%
%
%
\section{Conclusion}
\noindent In this work, we demonstrated the merits of bending beams for diverse applications in future wireless connectivity. We demonstrated that such beams are able to address the challenges of high frequency communications more efficiently than conventional beamforming, enabling both the adaptation of connectivity to the wireless environment, as well as the modification of the wireless environment according to the desired connectivity. We presented the design principles and capabilities of bending beams, we introduced the concept of near-field virtual routing and we proposed their potential application in use cases including user connectivity on curved trajectory, static/dynamic blockage avoidance, interference-free regions and multi-beam applications. The efficient formation of such beams using large arrays was analyzed with respect to crucial parameters, including the footprint size, the footprint shape, the array's inter-element spacing, the sub-array selection of active elements, the available phase levels and the operating frequency, giving promise for near-field wireless communications with user connectivity on curved trajectories.
\section*{Acknowledgments}
This work was supported by the European Commission’s Horizon Europe Programme under the Smart Networks and Services Joint Undertaking INSTINCT project (Grant Agreement $101139161$).
%
%
\appendix[ ]
\subsection{Input phase for propagation on curved trajectory}
\label{Sec:AppendixA}
To determine the input phase $\phi(x,y)$ that generates a desired curved trajectory, we start with defining the caustic $x_c=f(z_c)$, and we consider trajectories on the $xz$-plane (invariant along the $y$-direction). A caustic is the envelope of all rays that are launched from the aperture and each ray is tangent to the caustic. Hence, each point at the input transverse plane (at $z = 0$) can be functionally related to a point on the caustic via a slope of $\theta$, i.e. $\tan\theta$. Taking into account that the slope of the caustic at point ($x_c,z_c$), 
is $df(z_c)/dz_c$, we may write
\begin{align}
    \frac{df(z_c)}{dz_c}=\tan{\theta}=\frac{\sin{\theta}}{\sqrt{1-\sin^2{\theta}}},
    \label{Eq:EqA1}
\end{align}
or, solving in terms of $\sin{\theta}$
\begin{align}
    \sin{\theta}=\frac{df(z_c)/dz_c}{\sqrt{1+(df(z_c)/dz_c)^2}}.
    \label{Eq:EqA2}
\end{align}
The phase $\phi$ determines the in-plane $k$-vector of the wave and, hence, its derivative is associated with the local in-plane $k$-component as 
\begin{align}
    \frac{d\phi(x)}{dx}=k\sin{\theta}.
    \label{Eq:EqA3}
\end{align}
Combination of \eqref{Eq:EqA2} with \eqref{Eq:EqA3} leads to the final result
\begin{align}
    \frac{d\phi(x)}{dx}=k\frac{df(z_c)/dz_c}{\sqrt{1+(df(z_c)/dz_c)^2}}.
    \label{Eq:EqA4}
\end{align}
For paraxial rays we may approximate $\sin\theta\approx\tan\theta$, which leads to the simplified version of \eqref{Eq:EqA4} for the paraxial limit
\begin{align}
    \frac{d\phi(x)}{dx}=k\frac{df(z_c)}{dz_c}.
    \label{Eq:EqA5}
\end{align}
\subsection{Input phase for propagation on parabolic trajectory}
\label{Sec:AppendixB}
To integrate \eqref{Eq:EqPHASEnonpar} for a trajectory of the general form $x_c=f(z_c)$ we need to express the derivative $df(z_c)/dz_c$ in terms of the coordinates at the input $xy$-plane. Therefore, we need to associate each point of the caustic with the coordinates of a ray at the input plane. For beam bending that takes place on the $xz$-plane, a ray that is tangent to the caustic at point ($x_c,z_c$) is described analytically by
\begin{align}
    \frac{x-x_c}{z-z_c}=\frac{df(z_c)}{dz_c}.
    \label{Eq:EqB1}
\end{align}
At the input plane, where $z=0$, the ray equation reduces to
\begin{align}
    x=x_c-z_c\frac{df(z_c)}{dz_c},
    \label{Eq:EqB2}
\end{align}
which we can use to express $x_c,z_c$ in terms of $x$.\\
\indent For the parabolic trajectory of \eqref{Eq:EqCAUSTICAiry} we find that 
\begin{align}
    \frac{df(z_c)}{dz_c}=2\beta (z_c-z_0).
    \label{Eq:EqB3}
\end{align}
To solve \eqref{Eq:EqB2}, we replace $x_c$ and $df(z_c)/dz_c$ according to \eqref{Eq:EqCAUSTICAiry} and \eqref{Eq:EqB3}, respectively, and we obtain the result
\begin{align}
    z_c=\sqrt{\frac{x_0-x+z_0^2\beta}{\beta}}.
    \label{Eq:EqB4}
\end{align}
Hence, \eqref{Eq:EqB3} can be now expressed in terms of the coordinates at the input plane as
\begin{align}
    \frac{df(z_c)}{dz_c}=2\beta \left(\sqrt{\frac{x_0-x+z_0^2\beta}{\beta}}-z_0 \right),
    \label{Eq:EqB5}
\end{align}
i.e. we have explicitly expressed the r.h.s. of \eqref{Eq:EqPHASEnonpar} in terms of the specific trajectory given by \eqref{Eq:EqCAUSTICAiry}. We can now use \eqref{Eq:EqB5} to integrate \eqref{Eq:EqPHASEnonpar}. Integration leads to the generalized phase profile
\begin{align}
    \phi(x) = \frac{1}{4\beta}\left(-2(\Psi+4\beta z_0)\sqrt{1+4\beta\left(x_0-x-\Psi z_0\right)} + \Phi \right),
    \label{Eq:EqB6}
\end{align}
where $\Psi=2\sqrt{\beta(x_0-x+\beta z_0^2)}-2\beta z_0$ and $\Phi = \mathrm{arcsinh}\left(\Psi \right)$. For $x_0=z_0=0$, $\Psi\rightarrow2\sqrt{-\beta x}$ and \eqref{Eq:EqB6} reduces to \eqref{Eq:EqPHASEnonparPARABOLIC}. For slowly bending beams we may integrate \eqref{Eq:EqA3} instead of the full form of \eqref{Eq:EqPHASEnonpar}, which leads to the phase profile given in \eqref{Eq:EqAiryPHASE}.
\subsection{Numerical propagation in free-space}
\label{AppendixC}
Beams can be generally expressed as linear superposition of plane waves as
\begin{equation}
    E(x,y) = \iint dk_xdk_y \widetilde{E}(k_x,k_y) e^{j(k_xx+k_yy)},
    \label{Eq:EqX01}
\end{equation}
where ($k_x,k_y$) are the transverse wavenumbers, and the tilde denotes operation in $k$-space. Note that $\widetilde{E}(k_x,k_y)$ is simply the inverse Fourier transform of $E(x,y)$ and corresponds to the $k$-dependent weights of the individual plane waves. Each plane wave propagates acquiring a phase $e^{jk_zz}$, where $k_z=\sqrt{k^2_0-k^2_x-k^2_y}$, limiting the propagating $k$-components within the range $(k_x^2+k_y^2)/k_0^2<1$. Because the beam is equivalently described by the superposition of the individually propagated plane waves, the beam propagates in $k$-space as $\widetilde{E}(k_x,k_y)e^{jk_zz}$, i.e., the $k$-content of the beam acquires a global phase without undergoing other changes in magnitude and phase (except for the case of lossy atmosphere or turbulent conditions, which are cases beyond the scope of this work). \\
\indent Taking into account that \eqref{Eq:EqX01} is the inverse Fourier transform of the beam, beam propagation from $z$ to $z+\delta z$ can be written concisely as
\begin{equation}
    E(x,y,z+\delta z) = \mathcal{FT}\,^{-1}\{\,\mathcal{FT}\,[E(x,y,z)]e^{jk_z\delta z}\},
    \label{Eq:EqX02}
\end{equation}
where \(\mathcal{FT}\) denotes the Fourier transform and \(\mathcal{FT}\)\,$^{-1}$ its inverse. Due to linearity, if the $k$-content is known at a certain propagation step, the beam can be reconstructed anywhere in real space.
\subsection[]{Derivation of $z_\mathrm{max}$ as a function of the RIS size}
\label{Sec:AppendixD}
For the parabolic trajectory \eqref{Eq:EqCAUSTICAiry}, the ray equation \eqref{Eq:EqB1} is written using the slope \eqref{Eq:EqB3} as
\begin{align}
    \frac{x-x_c}{z-z_c}=2\beta (z_c-z_0).
    \label{Eq:EqC1}
\end{align}
The maximum distance, $z_\mathrm{max}$, is determined by the ray originating from the edge of the aperture, which is located at $x=-L_x,z=0$. Using the coordinates of the edge ray to solve \eqref{Eq:EqC1} in terms of $z$ we reach the result \eqref{Eq:EqZmax}, for which we have used \eqref{Eq:EqCAUSTICAiry} to express $x_c$ in terms of $z_c$. 
\subsection{Spatial bandwidth of Airy beam with exponential tapering}
\label{Sec:AppendixE}
The Fourier transform of the Airy beam given by \eqref{Eq:EqAiryTHEORY} can be calculated in closed form, and is given by
\begin{align}
    E(k_x) = \frac{1}{2(4\beta k^2)^{1/3}}\exp{\left(i \frac{(kx+ia)^3}{12\beta k^2}\right)}
    \label{Eq:EqD1}
\end{align}
and, hence, the power spectrum follows a Gaussian distribution of the form
\begin{align}
    |E(k_x)|^2 \propto \exp{\left(-\frac{a}{2\beta k^2}k_x^2 \right)}.
    \label{Eq:EqD2}
\end{align}
The FWHM extent of the Gaussian distribution given by \eqref{Eq:EqD2} is 
\begin{align}
    \delta k_\mathrm{FWHM} = 2k\sqrt{\ln2\frac{2\beta}{a}},
    \label{Eq:EqD3}
\end{align}
which is expressed in normalized $k$-units as given by \eqref{Eq:EqBW}.
\bibliographystyle{IEEEtran}
\bibliography{IEEEabrv,main}
\newpage

\vspace{11pt}


\vspace{11pt}

\vfill
\end{document}